\documentclass[sigconf]{acmart}

\usepackage{todonotes}

\usepackage{listings}
\usepackage[symbol]{footmisc}

\usepackage{tikz}
\usepackage{amsmath}
\usepackage{multirow}
\usepackage{booktabs}

\usepackage{caption}
\usepackage{subcaption}

\usepackage{xcolor}
\usepackage{pgfplots}
\usepackage{tikz}
\pgfplotsset{compat=1.8}
\pgfplotsset{
    width=\textwidth,
}
\usepackage{lipsum}
\usepackage{pgfplotstable}
\usetikzlibrary{pgfplots.groupplots}

\newcommand*\circled[1]{\tikz[baseline=(char.base)]{
            \node[shape=circle,fill,inner sep=1pt] (char) {\textcolor{white}{#1}};}}

\usepackage{tikz}

\definecolor{codegreen}{rgb}{0,0.6,0}
\definecolor{codegray}{rgb}{0.5,0.5,0.5}
\definecolor{codepurple}{rgb}{0.58,0,0.82}
\definecolor{mGreen}{rgb}{0,0.6,0}
\definecolor{mGray}{rgb}{0.5,0.5,0.5}
\definecolor{mPurple}{rgb}{0.58,0,0.82}
\definecolor{backcolour}{rgb}{0.95,0.95,0.92}

\definecolor{RYB1}{RGB}{80, 99, 42}
\definecolor{RYB2}{RGB}{215, 227, 191}
\definecolor{RYB3}{RGB}{198, 187, 174}
\definecolor{RYB4}{RGB}{146, 205, 220}
\definecolor{RYB5}{RGB}{238, 144, 34}
\definecolor{RYB6}{RGB}{142, 172, 59}

\pgfplotscreateplotcyclelist{colorbrewer-RYB}{
{RYB1!50!black,fill=RYB1},
{RYB2!50!black,fill=RYB2},
{RYB3!50!black,fill=RYB3},
{RYB4!50!black,fill=RYB4},
{RYB5!50!black,fill=RYB5},
{RYB6!50!black,fill=RYB6},
}

\usepackage{lscape} 
\usepackage{pdflscape}
\usepackage{afterpage}
\usepackage{capt-of}
\usepackage{float}
\usepackage{array,multirow,graphicx}
\usepackage[titletoc]{appendix}
\usepackage{caption}
\usepackage{subcaption}
\usepackage{soul}
\usepackage{pifont}

\definecolor{ggreen}{HTML}{2CC225}
\definecolor{yyellow}{HTML}{C2C80A}
\definecolor{bbrown}{HTML}{8e4603}

\newcommand{\cmark}{\ding{51}}%
\newcommand{\xmark}{\ding{55}}%

\definecolor{codegreen}{rgb}{0,0.6,0}
\definecolor{codegray}{rgb}{0.5,0.5,0.5}
\definecolor{codepurple}{rgb}{0.58,0,0.82}
\definecolor{mGreen}{rgb}{0,0.6,0}
\definecolor{mGray}{rgb}{0.5,0.5,0.5}
\definecolor{mPurple}{rgb}{0.58,0,0.82}
\definecolor{backcolour}{rgb}{0.95,0.95,0.92}

\lstdefinestyle{CStyle}{
    commentstyle=\color{mGreen},
    keywordstyle=\color{magenta},
    numberstyle=\tiny\color{mGray},
    stringstyle=\color{mPurple},
    basicstyle=\sffamily\footnotesize,
    frame=lrtb,
    breakatwhitespace=false,         
    breaklines=true,                 
    captionpos=b,                    
    keepspaces=true,                 
    numbers=left,                    
    numbersep=5pt,                  
    showspaces=false,                
    showstringspaces=false,
    showtabs=false,                  
    tabsize=2,
    language=C
}

\lstdefinestyle{CStyle1}{
    commentstyle=\color{mGreen},
    keywordstyle=\color{magenta},
    numberstyle=\tiny\color{mGray},
    stringstyle=\color{mPurple},
    basicstyle=\sffamily\footnotesize,    frame=lrtb,
    breakatwhitespace=false,         
    breaklines=true,                 
    captionpos=b,                    
    keepspaces=true,                 
    numbers=left,                    
    numbersep=5pt,                  
    showspaces=false,                
    showstringspaces=false,
    showtabs=false,                  
    tabsize=2,
    language=C
}

\lstdefinestyle{mystyle}{
    commentstyle=\color{codegreen},
    keywordstyle=\color{magenta},
    numberstyle=\tiny\color{codegray},
    stringstyle=\color{codepurple},
    basicstyle=\sffamily\footnotesize,
    breakatwhitespace=false,         
    breaklines=true,                 
    captionpos=b,                    
    keepspaces=true,                 
    numbers=left,                    
    numbersep=5pt,                  
    showspaces=false,                
    showstringspaces=false,
    showtabs=false,                  
    tabsize=2,
    language=C
}
\lstdefinestyle{trans}{
    commentstyle=\color{codegray},
    numberstyle=\tiny\color{codegray},
    stringstyle=\color{codepurple},
     basicstyle=\sffamily\footnotesize,
    frame=lrtb,
    breakatwhitespace=false,         
    breaklines=true,                 
    captionpos=b,                    
    keepspaces=true,                 
    numbers=left,                    
    numbersep=5pt,                  
    showspaces=false,                
    showstringspaces=false,
    showtabs=false,                  
    tabsize=2,
     language=[x86masm]Assembler,  escapeinside={\%*}{*)},   
     }     

\AtBeginDocument{%
  \providecommand\BibTeX{{%
    \normalfont B\kern-0.5em{\scshape i\kern-0.25em b}\kern-0.8em\TeX}}}

\acmConference[HyperDbg]{}{May,
  2022}{ArXiv}
\acmPrice{}
\acmISBN{}

\begin{document}

\title{\bf HyperDbg: Reinventing Hardware-Assisted Debugging\\
(Extended Version)}


\author{\normalsize{Mohammad Sina Karvandi$^{1,2}$, MohammadHossein Gholamrezaei$^{1,2}$, Saleh Khalaj Monfared$^{3,*}$,\\ Soroush Meghdadizanjani$^{4,*}$, Behrooz Abbassi$^{2}$, Ali Amini$^{2}$, Reza Mortazavi$^{5}$, Saeid Gorgin$^{1}$, Dara Rahmati$^{1}$,\\ Michael Schwarz$^{6}$}}

 \affiliation{$^{1}$ Institute For Research in Fundamental Sciences (IPM), $^{2}$ HyperDbg Org., $^{3}$ Worcester Polytechnic Institute (WPI),\\ $^{4}$ Stony Brook University, $^{5}$ Damghan University, $^{6}$ CISPA Helmholtz Center for Information Security  }

\renewcommand{\shortauthors}{Karvandi, et al.}

\begin{abstract}

Software analysis, debugging, and reverse engineering have a crucial impact in today's software industry.
Efficient and stealthy debuggers are especially relevant for malware analysis. 
However, existing debugging platforms fail to address a transparent, effective, and high-performance low-level debugger due to their detectable fingerprints, complexity, and implementation restrictions.

In this paper, we present \textsc{HyperDbg},\footnote[1]{Contribution of these authors performed while at Institute For Research in Fundamental Sciences (IPM).} a new hypervisor-assisted debugger for high-performance and stealthy debugging of user and kernel applications. 
To accomplish this, \textsc{HyperDbg} relies on state-of-the-art hardware features available in today's CPUs, such as VT-x and Extended Page Table (EPT). 
In contrast to other widely used existing debuggers, we design \textsc{HyperDbg} using a custom hypervisor, making it independent of OS functionality or API. 
We propose hardware-based instruction-level emulation and OS-level API hooking via extended page tables to increase the stealthiness. 
Our results of the dynamic analysis of 10,853 malware samples show that \textsc{HyperDbg}'s stealthiness allows debugging on average 22\% and 26\% more samples than \textit{WinDbg} and \textit{x64dbg}, respectively.
Moreover, in contrast to existing debuggers, \textsc{HyperDbg} is not detected by any of the 13 tested packers and protectors.
We improve the performance over other debuggers by deploying a VMX-compatible script engine, eliminating unnecessary context switches. 
Our experiment on three concrete debugging scenarios shows that compared to \textit{WinDbg} as the only kernel debugger, \textsc{HyperDbg} performs step-in, conditional breaks, and syscall recording, 2.98x, 1319x, and 2018x faster, respectively.
We finally show real-world applications, such as a 0-day analysis, structure reconstruction for reverse engineering, software performance analysis, and code-coverage analysis. 


\end{abstract}





\maketitle

\section{Introduction}

Debuggers are an essential element in software development and analysis that are actively employed by computer engineers to improve efficiency, detect security flaws, and fix bugs in software programs. Additionally, debuggers are also utilized as a valuable tool for software reverse engineering and malware analysis purposes. There has been a series of commercial and open-source debugging software offering convenient features to address such needs \cite{windbg,GDB,LLDB,x64,OllyDbg}. Given the outstanding growth in the sophistication and complexity of evasion and obfuscation methods, it is necessary to facilitate powerful debuggers to analyze, detect, and understand malware. 

Modern binary executables, armed with packing \cite{you2010malware}, evasion \cite{d2020dissection}, and hardware-assisted techniques \cite{ning2018hardware,kim2019disabling}, employ a series of methods that involve anti-virtualization \cite{apostolopoulos2021resurrecting}, anti-emulation\cite{lictua2018anti}, as well as side effects and footprint detection \cite{afianian2019malware} (e.g., call to specific OS APIs) to impede debugging.
Despite many valuable efforts for development of transparent and effective analysis methods in the community \cite{dinaburg2008ether,deng2013spider,yan2012v2e,quynh2010virt,zhang2016towards,ning2017ninja}, currently available debugging tools struggle to encounter modern protected programs and malware. 
These tools lack elaborate kernel-side components to offer deep scrutiny for reverse-engineering purposes. A comprehensive analysis of 4 million malware samples shows that 88\% are equipped with anti-reversing, and 81\% with anti-debugging or virtualization techniques \cite{branco2012scientific}.
Utilizing OS APIs \cite{chen2016advanced} or leveraging ring-0 options \cite{gao2012debugging} leads to artifacts and leakages that high-privilege malware can detect.

All these complications have recently attracted the attention of researchers to integrate the debugging infrastructure deeper into the hardware.
As a result, solutions based on bare metal \cite{zhang2016towards,kirat2011barebox,willems2012down}, hypervisor-level (VT-x) \cite{deng2013spider,lengyel2014scalability, willems2013hypervisor, fattori2010dynamic}, System Management Mode (SMM) \cite{zhang2016towards}, or even Intel Memory Management Engine (MME) \cite{ermolov2019intel} are used to minimize the leakage of the debugger's presence.
This increases the transparency of the debugger and thus its stealthiness. 
While these lower-level realization of debugging mechanisms increase the transparency surface, they suffer from huge performance degradation. 
Although sub-kernel deployment \cite{zhang2015using} of debugging, monitoring and software analysis tools can offer a powerful platform for such use cases such as analyzing evasive malware, previously-proposed sub-kernel debuggers fail to provide rich debugging functionality as they have been either discontinued \cite{softice,fattori2010dynamic}, developed for pure academic purposes \cite{fattori2010dynamic}, or have not been through thorough development and testing required for dealing with real-world applications and scenarios \cite{zhang2015using}. 
Moreover, the availability of the source code for such tools is still known to be a requirement in the community. 

In this paper, we propose \textsc{HyperDbg}, a hypervisor-based (ring -1) debugger designed to use modern hardware technologies to provide new features to the reverse-engineering community. 
It operates on top of Windows by virtualizing an already running system using Intel VT-x. 
As a primary goal, \textsc{HyperDbg} strives to be as stealthy and OS-independent as possible.
\textsc{HyperDbg} avoids using any operating-system APIs and software debugging mechanisms.
Instead, it extensively uses processor features such as Second Layer Page Table, i.e., Extended Page Tables (EPT), to monitor both the kernel and the user executions.

Avoiding OS-based debugging APIs increases the transparency against classic anti-debugging methods. 
Moreover, by directly relying on hardware feature, \textsc{HyperDbg} is hard to detect with time-delta methods that detect the presence of hypervisors, e.g., by detecting the overhead of traps into the hypervisor \cite{oyama2019does,mcgraw2000attacking}.
Such hardware-enabled features also allows \textsc{HyperDbg} to offer various state-of-the-art functions such as hidden hooks, which are as fast as current inline hooks but also offer stealth debugging. 
\textsc{HyperDbg} supports \textit{Hardware Debug Registers} simulation to break on read and write accesses to a specific location while remaining entirely invisible to both the OS kernel and the programs. 
Moreover, such hardware-assisted features make it possible for \textsc{HyperDbg} to eliminate limitations previously imposed by \textit{Hardware Debug Registers} in size and count \cite{zhang2016towards}.
We evaluate the transparency by extensive evaluation against anti-debugging, anti-virtualization, anti-hypervisor methods, and packer software. 
\textsc{HyperDbg} was not detected by any of the 13 tested packers and protectors.
No other existing debugger achieves this level of stealthiness, with debuggers being detected on average by 44\% of packers and protectors, with no debugger detected by less than 3. 
We demonstrate the applicability of transparent debugging on 10,853 malware samples. 
Our results show that \textsc{HyperDbg} successfully analyzes 22\% and 26\% more malware samples compared to \textit{WinDbg} and \textit{x64dbg} respectively. 
We also describe an existing 0-day vulnerability in Windows 10 kernel successfully analyzed by \textsc{HyperDbg}'s transparent mode, rediscovered during our experiments.

For high-performance debugging, \textsc{HyperDbg} uses a VMX-root-compatible script engine that executes the entire debugging functionality in the kernel mode, enabling complex debugging functionality. 
Our script engine eliminates any user to kernel-mode interaction, making any OS-level API obsolete while providing a huge debugging performance. 
We evaluate the improved debugging performance in three concrete debugging scenarios: stepping, conditional breaks, and syscall recording. 
Compared to the state-of-the-art debugger \textit{WinDbg}, \textsc{HyperDbg} is 2.98, 1319, and 2018 times faster, respectively.

We show that the unique design of \textsc{HyperDbg} enables use cases beyond classical debugging scenarios.
We describe how the proposed debugger enables transparent debugging of I/O devices, analyses performance of software, and provides means for code coverage usable for (kernel) fuzzing. 
Finally, our analysis of a Windows 10 0-day in a kernel-mode bootkit malware shows that \textsc{HyperDbg} is mature enough for real-world malware analysis. 

\paragraph{\textbf{Contributions.}} The contributions of this paper are as follows.
\begin{enumerate}

\item We present \textsc{HyperDbg}, a hypervisor-assisted debugger specialized for deep software analysis, reverse engineering, and fuzzing with a focus on stealthiness. 

\item We introduce a VMX-root-compatible script engine within \textsc{HyperDbg} that is orders of magnitude faster than state-of-the-art debuggers for common tasks. 

\item We demonstrate transparent debugging on 10,853 malware samples, showing that \textsc{HyperDbg} can analyze 22\%-26\% more malware samples than state-of-the-art debuggers. 

\item We describe multiple applications of \textsc{HyperDbg}, such as large-scale and fast malware analysis including a Windows 0-day analysis, code coverage in fuzzing, debugging of I/O devices, and software-performance measurements.

\end{enumerate}

\paragraph{\textbf{Availability}}
\textsc{HyperDbg} is fully open source and is available to foster the security research and software engineering: \\\url{https://github.com/HyperDbg/HyperDbg}. 

\paragraph{Outline} The remainder of this paper is organized as follows.
In Section~\ref{sec:background}, we provide required background information.
Section~\ref{sec:design} presents the design, and Section~\ref{sec:arch} the architecture of \textsc{HyperDbg}.
Section~\ref{sec:engine} introduces the script engine, and Section~\ref{sec:transparency} the transparency-mode of \textsc{HyperDbg}.
Section~\ref{sec:evaluation} provides the transparency and performance evaluations.
Section~\ref{sec:applications} describes additional use cases.
Section~\ref{sec:relatedwork} discusses related work, and Section~\ref{sec:conclusion} concludes the paper.

\section{Technical Background}\label{sec:background}

In this section, we survey the technical background knowledge to describe the design of the proposed debugger. We briefly review the structure and features of modern debuggers, hypervisors, and the main hardware capabilities provided by Intel, on top of which \textsc{HyperDbg} is implemented.

\subsection{Modern Debuggers}

Debugging is fundamentally defined as the process of examination and analysis of a software program to understand or locate the unsatisfying code snippets in terms of functionality, performance, or security flaw \cite{aggarwal2002debuggers,ko2008debugging}. To address the desired functionalities, a debugger should facilitate multiple mechanisms. Stepping through the source code or assembly, memory inspection and modification, as well as breakpoint definition are vital features in commodity debuggers. 
From the reverse engineering and malware analysis perspective, debuggers generally fall into two categories of user-mode and kernel-mode debuggers \cite{gao2012debugging}. User-mode debuggers provide the basic functionality to analyze a user-mode process. They are simply implemented and easy to use. User-mode debuggers give a convenient and isolated environment for the user. \textit{x64dbg} \cite{x64}, \textit{Ollydbg} \cite{OllyDbg}, and Immunity Debugger \cite{immunitydbg} are well-known examples of user-mode debuggers.
Kernel-level debuggers run in kernel mode, which grants them higher privileges in terms of register and memory access during the program's execution. \textit{WinDbg} \cite{windbg} and \textit{GDB} \cite{GDB} are famous examples of kernel debuggers that are widely used for reverse engineering and malware analysis \cite{afianian2019malware}.
With advances in malware evasion techniques \cite{apostolopoulos2021resurrecting}, researchers have been showing interest towards virtualization, simulation, and hardware-assisted debugging methods \cite{leon2021hypervisor} that can offer a more transparent environment for code analysis and low-level modification of the execution flow \cite{zhang2016towards,ning2017ninja}.

\subsection{Instruction Set Architecture (ISA) Extensions}

In this section, we briefly describe Intel VT-x, Intel EPT, and Intel TSX ISA extensions employed in the proposed hypervisor-level debugger. 
Note that \textsc{HyperDbg} in its current format only supports Intel processors and is built based on Intel technologies and terminology. However, similar hardware features exist both for AMD and ARM processors that can be exploited likewise. Further description is provided in Appendix \ref{appen:amd}.

\paragraph{Intel Virtualization Technology (VT-x)}
Intel VT-x (formerly known as Vanderpool) is the hardware virtualization technology provided by Intel for IA-32 processors to simplify virtualization and increase the performance of VMMs \cite{neiger2006intel}. VT-x introduces new data structures and instructions to the ISA \cite{ferreira2008intel} and enables processors to act as if there were several independent processors to allow multiple operating systems to run simultaneously on the same machine. 

\paragraph{Intel Extended Page Table (EPT)}

Intel VT-x technology comes with a hardware-assisted Memory Management Unit (MMU) and the implementation of Second Level Address Translation (SLAT), known as Extended Page Table (EPT). By translating the Guest Physical Address (GPA) to Host Physical Address (HPA) on the CPU level \cite{VmwareIntelEptPerformance}, EPT eliminates the overhead associated with software-managed shadow page tables \cite{HardwareAssistedMMU}. In Intel’s design, each CPU core can use a separate EPT Table, which allows for multiple independent accesses from different OSs concurrently.

 \begin{figure*}[!t] 
\centering
\includegraphics[width=0.95\linewidth]{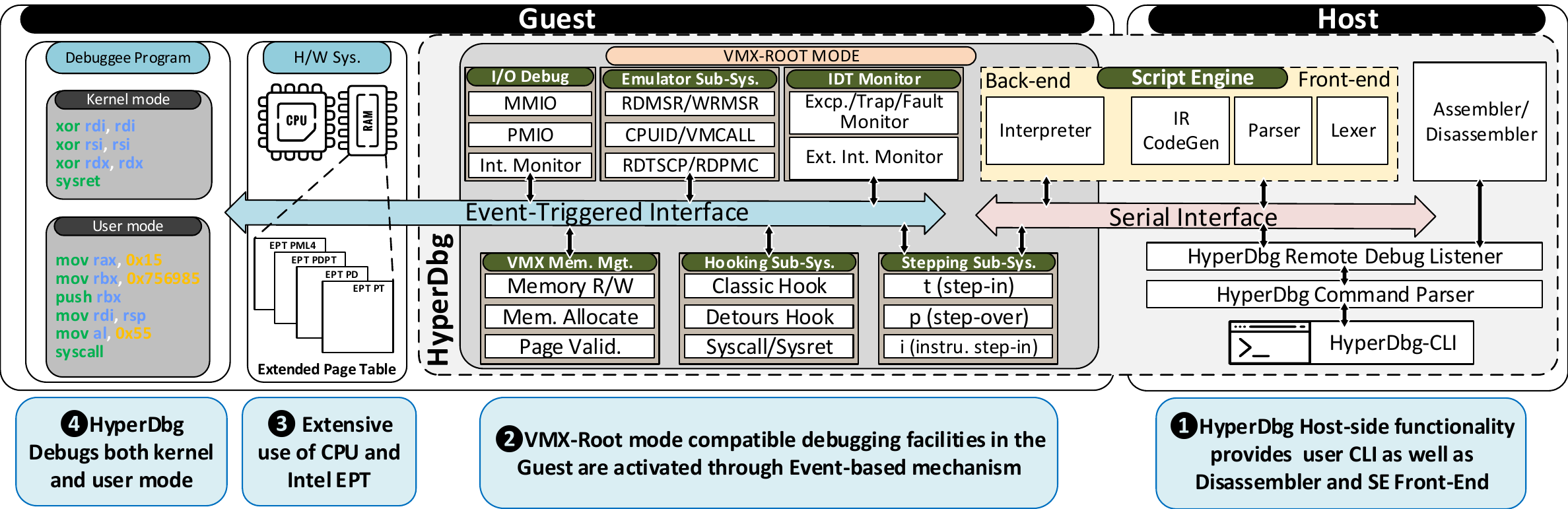}
\caption{High-level overview of \textsc{HyperDbg}'s sub-systems and execution flow}
\label{overview}
\end{figure*}

\paragraph{Intel Transactional Synchronization Extensions (TSX)}


Intel TSX is the product name for a set of x86 instruction set extensions, including Restricted Transactional Memory (RTM), which adds support for the declaration of hardware transactions. 
Instructions inside such a transaction  either all succeed or are rolled back altogether, in case any error or abortion occurs during the transaction, e.g., interrupt from OS~\cite{stecklina2018lazyfp, karvandi2022tsx}. In this paper, by using the term Intel TSX, we refer to RTM specifically.


\subsection{Terminology}
As the implementation here is based on \textit{Intel processors} and the target OS is \textit{Microsoft Windows}, we describe the low-level design of our system based on Intel and Windows terminologies. 

\paragraph{Hypervisor}

A hypervisor (also known as a virtual machine monitor or VMM) is a software that makes virtualization possible by virtually sharing the resources, such as memory and processor \cite{VmwareHypervisor,IbmHypervisor,PhoenixnapHypervisor}. It abstracts guest machines and the operating system from the actual hardware and runs virtual machines (VMs).




\paragraph{Interrupt Request Level (IRQL)}

An Interrupt Request Level (IRQL) is a hardware-independent mechanism that Windows uses to prioritize interrupts and code. Processes running at a higher IRQL preempt a thread or interrupt running at a lower IRQL.~\cite{IrqlMicrosoft}. 



There are different IRQLs used for different things. DIRQL is for interrupt service routines (ISRs) of hardware and external devices, DISPATCH\_LEVEL is used for the scheduler, DPCs, and code protected by spinlocks, APC\_LEVEL is for asynchronous procedure call (APC) routines, and PASSIVE\_LEVEL for user code, dispatch routines, and PnP routines.
 Appendix \ref{appen:C} gives a complete description of the used terms featured by Intel, which are also briefly tabulated in Table \ref{tab:term}.

\section{High-Level Overview}\label{sec:design}

This section provides a brief  high-level description of the design of \textsc{HyperDbg} and its building blocks. Here, we describe how the proposed debugging functionalities are implemented by a high-level abstraction and propose three debugging operations modes.


\subsection{High-level Debugging Flow}

On the high level, like other debuggers, \textsc{HyperDbg}  is designed to perform a level of analysis within a target system referred as the \textit{Guest}. The source debugging instructions are usually sent from an external system known as the debugger \textit{Host}. Figure \ref{overview} illustrates a high-level overview of \textsc{HyperDbg}'s sub-systems and execution flow. As shown, the debugger is an end-to-end framework, connecting the guest and the host systems by a communication interface (e.g., Serial). While the core building blocks are all deployed within the hypervisor-level on the guest side, the host side provides a CLI interface with the user and deploys an assembler/dissembler as well as a front-end engine for the debugging functionalities.
Multiple debugging sub-systems are deployed in the VMX-root mode of the guest system, which directly utilizes hardware features (e.g., EPT) for their functionality. As shown in Figure~\ref{overview}, the debugging commands are taken by the host where they are (dis)assembled and parsed through the script engine in \circled{1}. Then, the commands are sent via a communication channel to the guest. These commands are interpreted on the script engine's back-end at the guest's hypervisor-level. Based on the requested debugging routine, any user or kernel-mode debuggee program code can be targeted on the guest side with a direct access to the execution flow as indicated in \circled{4}. The sequence of the commands and functionalities are executed based on an event-triggered routine (Section \ref{sec:event}) according to each sub-system as depicted in \circled{2}. Finally, The sub-system functionalities utilize hardware-based features (e.g., EPT) to execute their operation in \circled{3}. We describe the deployment of each sub-system in detail in Section \ref{sec:arch}. Note that in the practical debugging procedure, bi-directional communication is required between the host and guest. However, as shown in the figure, with the use of the script engine, \textsc{HyperDbg} can confine the communication in an automated routine within the guest kernel mode if necessary.

\subsection{Event-Triggered Interface}
\label{sec:event}
To facilitate the debugging routines, we control the usage of the underlying functions and building blocks by an abstracted concept referred as an \textit{Event} in \textsc{HyperDbg}. Subsequently, we define \textit{Conditions} and \textit{Actions} that are used in the sub-system procedures for debugging.

\subsubsection{Events}

 An \textit{Event} is the occurrence of an incident that is of interest to the debugger. This comprises a wide range of activities ranging from a specific system call (\textit{Syscall}) that the debugger is set to monitor, to access to a particular memory address. \textsc{HyperDbg} can be configured to perform arbitrarily defined actions upon the occurrence of each event. A list of the supported events provided by \textsc{HyperDbg} is presented in Table \ref{tab:events} in the Appendix \ref{appen:A}.

\subsubsection{Actions}

Upon having an event triggered, \textsc{HyperDbg} can invoke specific functionalities known as actions. \textsc{HyperDbg} provides three types of action: \textit{Break}, \textit{Script}, and \textit{Custom Codes}. 
The Break action is the conventional feature of classic debuggers where all processing cores are paused until the debugger’s further permission.
The Script action allows viewing and modifying parameters, registers, and memory contents without breaking into the debugger. It also permits creating logs and running codes in the kernel space.
The Custom Codes action provides the ability to run custom assembly codes whenever a specific event is triggered.

\subsubsection{Conditions}

Conditions are specific circumstances that can be defined by the user in form of logical expressions to constrain the execution of an event. This, in turn, allows for the definition of conditional events where an event is triggered only upon evaluation of an expression to true.

\subsection{Operating Modes}

Based on different applicability, \textsc{HyperDbg} provides two modes of operation described as follows.
\subsubsection{VMI Mode}
Virtual Machine Introspection (VMI) Mode is presented for regular user application debugging and kernel-mode local debugging. Although it offers a conventional debugging experience by providing access to all \textsc{HyperDbg} features (including debugging, halting, and stepping user-mode applications) in an out-of-the-box fashion, kernel-mode breaking to the debugger and stepping are limited. VMI mode also allows scripts and custom codes in both user-mode and kernel-mode for local or remote debugging.
\subsubsection{Debugger Mode}

Debugger Mode is a powerful operating mode that allows for connecting to the kernel and halting the system to step-in and step-over through the kernel and user instructions. Here, debugging connectivity is carried out with a serial cable or a virtual serial device.



\subsubsection{Transparent Mode}

Both modes can be used in Transparent Mode, which offers stealth debugging by attempting to conceal \textsc{HyperDbg}’s presence on timing and micro-architectural levels. While the adversarial dynamic between malware generators and anti-malware producers is a never-ending process and this mode does not guarantee 100\% transparency, it makes it substantially more challenging for the anti-debugging and anti-hypervisor methods to detect the debugger. It is noteworthy to mention that \textsc{HyperDbg} is already immune to high-level anti-debugging methods that rely on API-specific methods to detect debugging environments (e.g., self-debugging binaries). The presenting transparency methodology is described in Section \ref{sec:trans} and is thoroughly evaluated in Section \ref{sec:transeval}. 









\section{Back-End Architecture}
\label{sec:arch}

This section explores the architectural design of \textsc{HyperDbg} on a sub-system level. We describe the challenges and shortcomings of the existing methods and debuggers for each sub-system. Then, by describing the underlying detailed  implementation of the core sub-systems, we propose \textsc{HyperDbg}'s approach to address each of these challenges. 

\subsection{Stepping Subsystem}

In this section, we investigate the stepping mechanism used in conventional debuggers and their shortcomings with regards to their capability in delivering a true line-by-line stepping procedure. We discuss the solutions offered in \textsc{HyperDbg} as a \textit{VMX-root} mode debugger to provide different stepping mechanisms and address these issues.

\subsubsection{Step-in}

Step-in offers the conventional step functionality available in commodity debuggers (e.g., \textit{WinDbg}~\cite{windbg}, \textit{GDB}~\cite{GDB}) by setting the \textit{RFLAGS} trap flag to make the system stop after execution of a single instruction. This allows the debugger to read/modify the content of the registers and the memory by following a trap flag in the kernel. 

\textbf{Challenge.} Conventional stepping mechanisms cannot guarantee a line-by-line stepping procedure as all other CPU cores and processes may execute their routines, and interrupts can drastically alter a program's execution flow.

\begin{figure*}[t]
     \centering
     \begin{subfigure}[b]{0.3\textwidth}
         \centering
         \includegraphics[width=\textwidth]{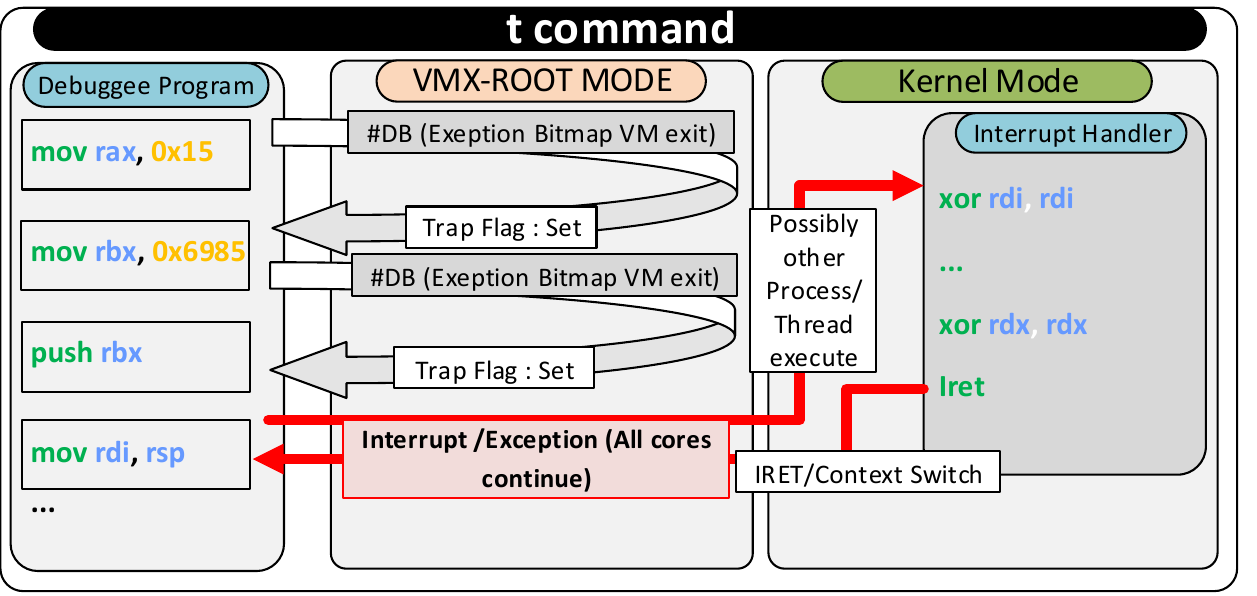}
\caption{The \textit{t command} Stepping mechanism in \textsc{HyperDbg}}
\label{step-t}
     \end{subfigure}
     \centering
     \begin{subfigure}[b]{0.35\textwidth}
    \centering
\includegraphics[width=1\textwidth]{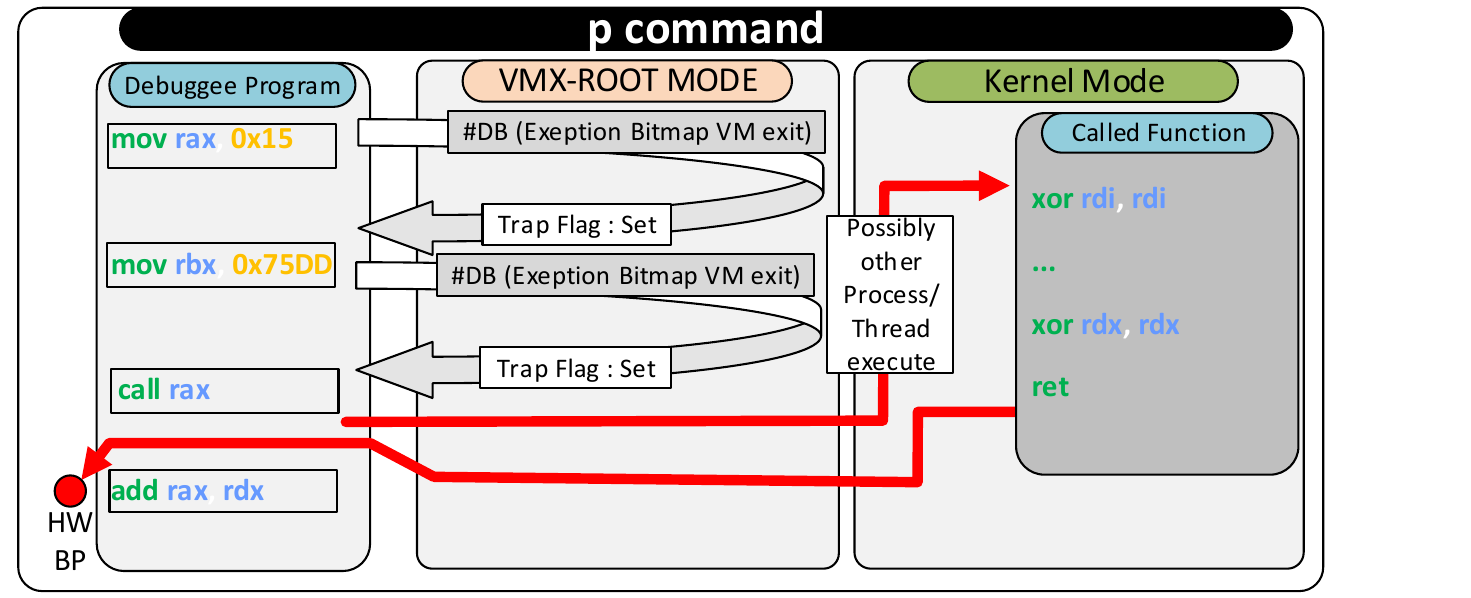}
\caption{The \textit{p command} Stepping over mechanism in \textsc{HyperDbg}}
\label{step-p}
     \end{subfigure}
     \begin{subfigure}[b]{0.3\textwidth}
        \centering
\includegraphics[width=1\textwidth]{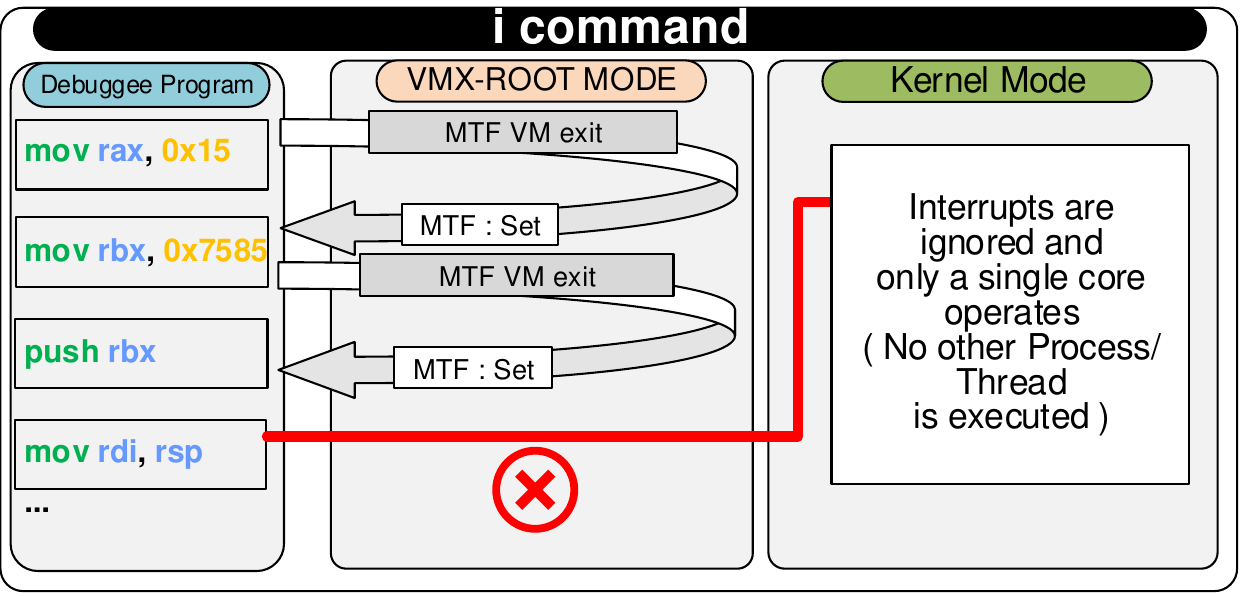}
\caption{The \textit{i command} Instrumentation Stepping Approach in \textsc{HyperDbg}}
\label{step-i}
     \end{subfigure}
        \caption{\textsc{HyperDbg} Stepping Commands}
        \label{fig:three graphs}
\end{figure*}

Figure \ref{step-t} shows an example of the step-in where the execution flow is disrupted by a \#DB exception interruption. A naive solution would mask all external interrupts by clearing the Interrupt Flag in \textit{RFLAGS}. However, intercepting/preventing the interrupts can easily break the OS semantics.
\textsc{HyperDbg} introduces the \textit{instrumental step-in} to provide a guaranteed stepping mechanism in debugging routine.

\textbf{Approach. }Considering the shortcomings of the conventional Step-in mechanism, \textsc{HyperDbg} introduces an instrumentation Step-in mechanism by employing the Monitor Trap Flag (MTF); a feature that works similar to RFLAGS's Trap Flag (TF) but appears transparent to the guest. Moreover, Non-Maskable Interrupts (NMIs) are used to ensure that the execution is done on a single core while other cores are halted. This method entirely overcomes the disruptions by inevitable interrupts.

\subsubsection{Instrumentation Step-in}

{\begin{sloppypar}
To the best of our knowledge, \textsc{HyperDbg} is the first debugger to address the issue by presenting a guaranteed stepping method. According to Figure \ref{step-i}, after executing the target instruction, a VM-exit is triggered (as an MTF has been previously set). Doing so guarantees that only the succeeding instruction is executed in the debugging guest. To do so, \textsc{HyperDbg} continues on only one core and disables interrupts on the same core (ignoring external interrupts by setting the external-interrupts exiting bit in VMCS) to offer a fine-grained stepping. This method provides the user with the unique feature to instrument routines from user-mode to kernel-mode and kernel-mode to user-mode that is not possible though other kernel debugger (\textit{WinDbg}). As an example, whenever the user-mode executes a SYSCALL instruction, \textsc{HyperDbg} allows the user to follow the instructions directly into the kernel and execute the next instruction in the kernel-mode (SYSCALL handler). Similarly, if a page-fault occurs in the middle of a user-mode application, the debugger is moved into the kernel-mode's page-fault handler. Kernel-mode to user-mode migration is also handled by \textsc{HyperDbg}, e.g., executing a \textit{SYSRET} or \textit{IRET} returns the debugger to user-mode from kernel-mode. 
\end{sloppypar}}
\subsubsection{Step-over}
The step-over mechanism in \textsc{HyperDbg} is very similar to conventional Step-in, except for the call instruction where the debugger sends the length of the call instruction to the debuggee, and instead of setting the Trap flag, it sets a Hardware Debug Register to the instruction after the call. Therefore, when the call is finished, the Hardware Debug Register is triggered, and the debugger is notified about the next instruction. Since other threads/cores might also trigger the Hardware Debug Register (as all the threads/cores are continued through the stepping), \textsc{HyperDbg} ignores such \#DBs from other Thread IDs/Process IDs and re-sets the debug register until reaching the correct execution context and target thread that is supposed to trigger the Hardware Debug Register. Figure \ref{step-p} shows the overview of the step-over stepping mechanism in \textsc{HyperDbg}, where upon inspection of a call instruction, a debug breakpoint exception (\#DB) is thrown for the next instruction.



\subsection{Hooking Subsystem}

Hooking in the context of debugging is the act of intercepting an arbitrary event (e.g. execution of a breakpoint on a particular address), running specific commands, and turning the execution flow back to the conventional routine at the entry point of the event. 

\textbf{Challenge.} Existing hooking systems in commodity debuggers implement direct memory access, which a user-mode software can easily check and detect. The integrity of memory can effortlessly be verified as well. This leaves the possibility of debugging detection for evasive malware. Moreover, \textit{Hardware Debug Registers} used to record memory content in debugging process are fixed in number and size, limiting hooking performance.

\subsubsection{SYSCALL and SYSRET Hooks}

\textsc{HyperDbg} implements hooking functionality by triggering an undefined opcode exception (\#UD) (by clearing the SCE bit in the Extended Feature Enable Register, i.e., \textit{IA32\_EFER}) and checking for the originating cause of the exception. The user can execute arbitrary scripts and set hooks for arbitrary system calls through the OS (\textit{SYSCALL}) or any return of the execution flow from a system-call (\textit{SYSRET}). During a user-to-kernel or kernel-to-user emulation, the debugger can monitor, execute or modify the system context before the actual execution of the instructions. \textsc{HyperDbg} provides the following approach for its novel hooking capabilities.

\textbf{Approach. }\textsc{HyperDbg} allows the user to monitor and manipulate memory accesses while remaining transparent by providing two EPT hooking mechanisms that reveal an unmodified version of the target page to the application. This methodology delivers an entirely transparent memory hook via EPT. Furthermore, we emulate Debug Registers to increase address traceability surpassing the previous limitations.  

\subsubsection{EPT Hidden Hook}
We propose EPT-level hooks that are not visible to the user-mode program or the operating system when attempting to read the hooking address. The first type of hidden breakpoints in \textsc{HyperDbg} are Classic EPT Hooks, which are achieved by injecting a \#BP (0xcc) to the target machine’s memory to cause a trap upon an attempt from the guest to execute the target memory address. The second variation of hidden hooks utilizes Detours-Style Hooks \cite{brubacher1999detours} (Inline EPT Hooks), which change the execution path by jumping to the patched instructions and returning the execution flow to the regular routine after the callback. While the latter approach has some flexibility constraints (e.g., limitations with the usage of the script engine, the range of hookable addresses, number of hooks in a page table), avoiding the costly VM-exit operation makes for a substantially faster hooking mechanism.


\subsubsection{Limitless Simulating of Debug Register (monitor)}

EPT hooking also allows for monitoring any read/write to any range of addresses by causing an event trigger to emulate \textit{Hardware Debug Registers} capability while eliminating its limitations on the number and lengths of trackable addresses \cite{chiueh2008fast}.




\subsection{Memory Access in VMX-root Mode}

Implementation of safe memory access is one of the challenging parts of designing a hypervisor-level debugger, as there are many scenarios that can lead to system halt or an exception (e.g., access to paged-out \cite{hand1999self} pages in the VMX-root \cite{SinaKarvandi2019HVFS_Part7}, and access to user-space memory from the VMX-root mode) that cannot be addressed using readily-available primitive instructions (e.g., \textit{mov}). 

\textbf{Challenge.} Safe memory access through VMX-level is extremely complicated as it is often handled by the OS. This often results in performance overheads and footprints in conventional current debuggers. However, safe and efficient memory access is necessary for many use cases such as malware analysis.


\textbf{Approach.} We propose a series of methodologies to address the complications of VMX memory management, described as follows. 

\subsubsection{ Discovering Page-table Entries}
The conventional method in \textsc{HyperDbg} to detect a valid page is checking for the presence of a valid page-table entry (with set present bit) for its target address. This method requires traversing through the page tables to carry out the discovery process. As an alternative method, we make use of Intel TSX. TSX suppresses exceptions/faults without any switch between user/kernel modes. 
This ability is leveraged in \textsc{HyperDbg} to check for the validity of a page by checking the successful execution of a transaction involving the target address. A similar approach has been used by Schwarz et al. \cite{schwarz2019practical} to check if an address in SGX is mapped. This method can be carried out using only a few instructions (Listing \ref{lst:exemplo}); however, as not all processors support this capability, \textsc{HyperDbg} automatically checks for the processor's support of this feature and switches to the former method if necessary. Our experiments show that a TSX-based page discovery for user-mode debugging is roughly three orders of magnitude faster since normal traversing requires the requests to be forwarded to the user for validity check. However, in  kernel-mode applications, the method incurs a 40\% slow-down due to the domination of cycles introduced by RTM routines.

\begin{lstlisting}[caption={Using Intel TSX to detect address validity.}, label={lst:exemplo},style=trans]
 ; Use Intel TSX to suppress any
; page-fault in VMX-root mode 
 %*\color{magenta}XBEGIN*)  $+xxx   ; End of TSX
 %*\color{magenta}MOV*) RAX, Dword PTR:[RCX] 
; Access the target memory address,
 %*\color{magenta}XEND *)   ; End of TSX
 %*\color{magenta}MOV*) RAX, 1
 %*\color{magenta}JMP*) Return
 %*\color{magenta}MOV*) RAX, 0
Return :
 %*\color{magenta}RETN *)   ; Return the result 

\end{lstlisting}

\subsubsection{Retrieving a Page by Injecting Page Fault (\#PF)}

Upon absence of a page, \textsc{HyperDbg} injects a page-fault to the debuggee (by configuring the \textit{CR2} register to the target virtual address) to request the VMX non-root to bring the page back from the hard disk to the RAM when it is resumed. While this method is not applicable in some scenarios (e.g., in DISPATCH\_LEVEL IRQL level as paging is not available), it can be useful in many others (e.g., upon execution of a SYSCALL or SYSRET where the system is guaranteed to be in PASSIVE\_LEVEL). 

\subsubsection{VMX-root Mode Compatible Message Tracing}\label{messagetrace}
Sending a message from VMX-root mode to VMX non-root mode is a challenging part of hypervisor design due to various limitations of accessing paged-pool buffers in VMX-root mode. Notably, most NT functions are not \textit{ANY IRQL} compatible, as they might access buffers that reside in paged pool memory. To send commands and messages from VMX-root mode to the user-mode application or the debugger, \textsc{HyperDbg} provides a custom VMX-root mode compatible message tracing mechanism.  This mechanism operates on the non-paged pool, and its memory is visible in VMX-root mode. By deploying specialized messaging buffers, we ensure that the messages are only sent when the paging process is safely accessible on the kernel-mode. The details of this mechanism is thoroughly discussed in \cite{hv8}.
\subsubsection{Reading and Writing Memory.}

Due to the various safety considerations surrounding making direct access to a user-space address from VMX-root mode, \textsc{HyperDbg} is designed not to access the memory directly but to use a virtual addressing method to reserve a page-table entry and map the desired user-mode physical address to a kernel-mode virtual address to enable safe memory read/write access. Furthermore, the write-enable bit in the PTE eliminates the check for the writability of the target address.

\subsubsection{Pre-allocated Pools.} 

Given that most of \textsc{HyperDbg}'s routines operate in VMX-root mode, \textsc{HyperDbg} makes use of pre-allocated pools to provide a mechanism for addressing the conventionally impossible \cite{SinaKarvandi2019HVFS_Part7} issue of allocating memory in the VMX-root mode. These pools (when divided into 4KB granularity) provide the resources necessary for EPT hooks. \textsc{HyperDbg}'s memory manager routines periodically check for any deallocation/replacement of memory pools needed in VMX root mode and performs them when the debugee is in VMX non-root mode.

\section{Front-End Architecture}
In the following section, we explore the intermediatory components of \textsc{HyperDbg}'s connecting back-end VMX-root mode sub-systems with the host machine as well as the user-interface functionality. Specifically, We describe guest-host communication and the kernel-level script engine. Although the core functionality of the proposed script engine operates on the guest side's VMX-root, we regard all non-VMX-root sub-modules in our framework as front-end here.
\subsection{Communicating and Task Appliance}

\begin{figure}[t] 
\centering
\includegraphics[width=0.9\linewidth]{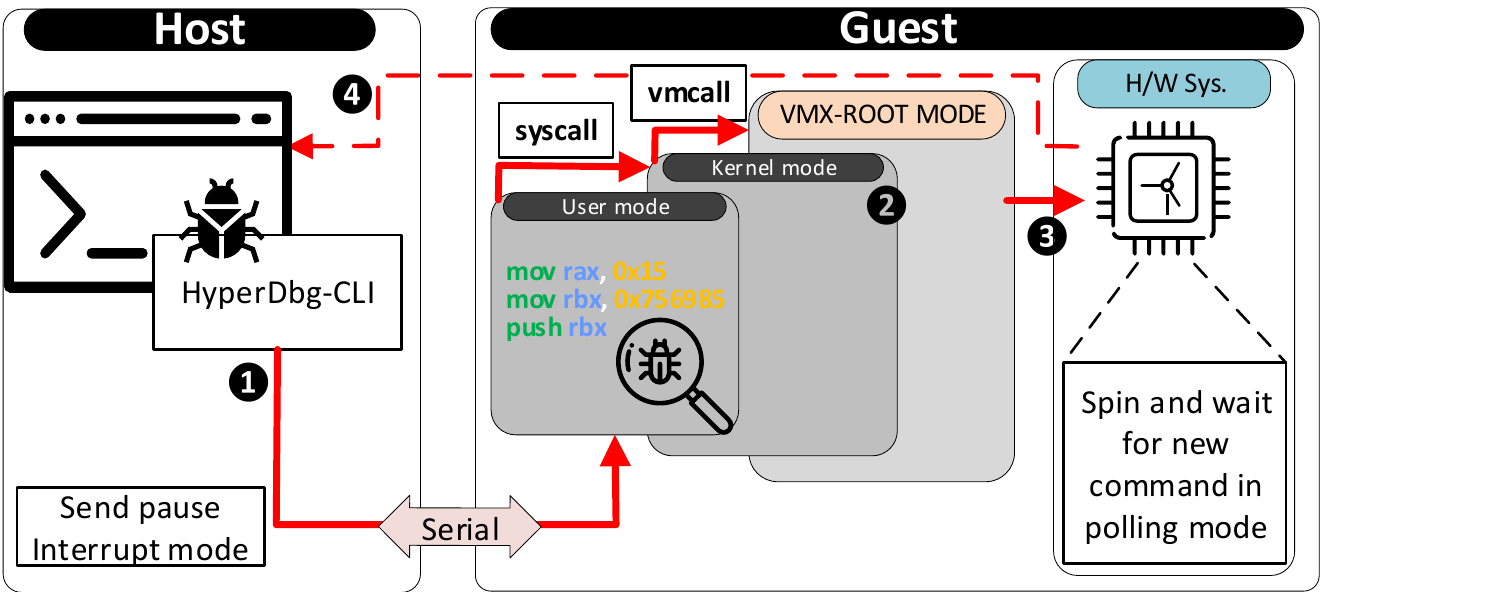}
\caption{The overall view of the communication in \textsc{HyperDbg}}

\label{com}
\end{figure}

The impracticality of using Windows API for data transmission over network in a debugger can be attributed to the unavailability of interrupts in VMX-root mode (which forces the mode of communication to polling mode) and the need for extra implementation, as Windows uses different device stacks in different IRQL levels for networking.
Owing to these challenges, \textsc{HyperDbg} utilizes serial ports for data transfers as it simplifies many aspects of design and usability and enables the use of polling mode.  Figure \ref{com} shows the general overview of \textsc{HyperDbg} communication routine. 
In addition to the serial communication, which is the conventional mode of communication in \textsc{HyperDbg}, KDNet functionality \cite{KDNet}, which is publicly available from Windows SDK \cite{winsdk}, has also been incorporated in \textsc{HyperDbg} as an alternative and more modern means of communication.

\subsubsection{Sending Data over Serial}
 
Following the connection initialization between a serial device and its corresponding serial port, a connection to the target device can be established by providing the COM argument. \textsc{HyperDbg} supports up to four different serial ports at a time.
Furthermore, halting a debuggee is performed by sending an interrupt signal using the interrupt mode of the serial device, which eliminates the need for gritty checks in polling mode when the debuggee is running. The interrupt to the user-mode application of the debuggee is passed down into the kernel-mode, where eventually, a VMCALL is invoked to put the debuggee to the pause state in the VMX-root mode and await further commands (packets) from the debugger.

\subsubsection{Communication between Cores}

Upon an event getting triggered, \textsc{HyperDbg} checks for a corresponding action and halts every other core in the \textit{VMX-root} mode in case of a break action (by sending Non-Maskable Interrupts (NMIs)~\cite{corporporation2018intel}, which cause the core to spin on a spinlock and invoke a \textit{VM-exit} and await further commands from the debugger), or executing the custom code/script without notifying the other cores, otherwise.

\subsection{Kernel-level Script Engine}\label{sec:engine}
\begin{figure}[t]
  \begin{subfigure}[b]{0.48\columnwidth}
    \includegraphics[width=\linewidth]{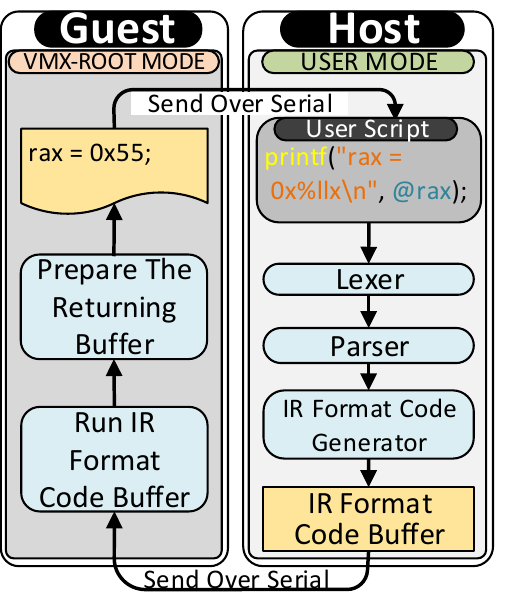}
\caption{\textsc{HyperDbg}s script engine's execution flow}

\label{se3}
  \end{subfigure}
  \hfill 
  \begin{subfigure}[b]{0.5\columnwidth}
    \includegraphics[width=\linewidth]{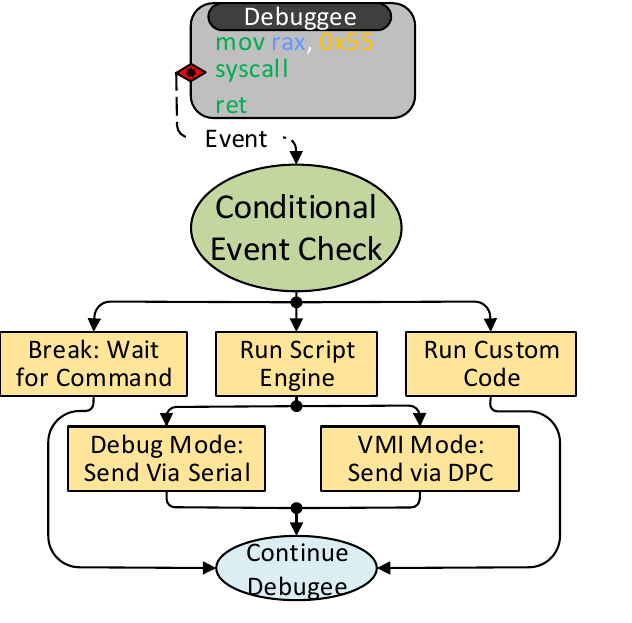}
\caption{Script Engine invocation as an \textit{event}}

\label{se1}
  \end{subfigure}
  \caption{ Description of Script Engine in \textsc{HyperDbg}  }
\end{figure}

Modern day debuggers fall short in providing a high-performance and highly customizable scripting framework. Striving to address this gap and faced with the lack of support for direct access to memory in VMX-root mode, we designed a VMX-enabled script engine from scratch. To the best of our knowledge, this is the only script-engine solution available in VMX-root mode offering advantageous features like OS spinlock, memory check as well as auxiliary functions (e.g., \textit{printf} and \textit{strlen}). As shown in the overview of the script engine's architecture in Figure \ref{se3}, the script engine is comprised of a back-end (that uses \textit{LL(1)} and \textit{LALR(1)} parsers for maximum efficiency) and a front-end that uses a \textit{MASM Style} syntax with C keywords (e.g., if, else, for) and an easily customizable grammar. 

The user-inputted scripts are delivered to the front-end host, scanned via a lexer, and parsed into an Intermediate Representation (IR), which is sent into a buffer over the serial interface into the guest's kernel VMX-root mode for execution. Afterward, a buffer is gradually filled with the execution results and transmitted back to the host. This approach offers substantial performance improvement compared to the conventional bidirectional method used in commodity debuggers (where commands and scripts are sent and parsed line by line) by sending the entirety of the script into the VMX-root mode, and the response back into the user mode, in a unidirectional flow.

As illustrated in Figure \ref{se1}, it is also possible to set a script as the \textit{action} of an \textit{event}. In this scenario, the parsed IR script is stored into the VMX-root kernel once, and upon having its corresponding event triggered, the IR is performed locally, thus improving the execution performance of the script engine. A sample script with a detailed description of the example is provided in Appendix~\ref{scriptenginecodesnippet}.

\section{Transparency Analysis}\label{sec:transparency}
\label{sec:trans}
In this section, we investigate the side effects and overhead created by \textsc{HyperDbg} which potentially could be exploited for detection. We further analyze different levels of transparency analysis using malware anti-debugging methods. Furthermore, we propose a statistical approach for hardening \textsc{HyperDbg} against timing side-channel attacks targeting sub-OS intercepting entities.\\

\subsection{Hypervisor Detection Methods and Mitigations}

Detection of sub-OS third-party programs (e.g., hypervisors) is carried out by querying for a set of indicative footprints, such as registry keys, system-calls (e.g., to discover running processes and loaded drivers), and instructions \cite{191958} (e.g., CPUID, IDT, LDT). \textsc{HyperDbg} counters these endeavors by intercepting the attempt, forcing a VM-exit, and emulating the corresponding return values with those of a normal, non-virtualized environment in the VM-exit handler. Table 1 provides a comprehensive overview of these methods.
More sophisticated hypervisor/VM detection methods exploit timing side channels. The key idea is the fact that certain instructions (e.g., CPUID, GETSEC, INVD, XSETB) cause a VM-exit routine when executed. If the target program is running in a VM, this results in a longer execution time than on bare metal, which can be detected by timing measurements. Listing \ref{lst:exemplo1} shows an example of such attacks. In the following, we describe the mechanisms in \textsc{HyperDbg} to counter these detection methods.

\begin{table}[b]
\centering
\caption{Anti-Debugging and Anti-VM exercises and mitigation in \textsc{HyperDbg}}
\resizebox{1\columnwidth}{!}{%
\begin{tabular}{ccccc}
\hline
\textbf{Cat.}                                                   & \textbf{Methodology}                                                             & \textbf{Example of the Meth.}                                                                                                                               & \textbf{Example} & \textbf{Mitigation in HyperDbg}                                                                                                             \\ \hline
\multicolumn{1}{c|}{\multirow{7}{*}{{\rotatebox[origin=c]{90}{\centering \textit{Anti-Debugging and Fingerprinting Methods}}}} } 
              & \begin{tabular}[c]{@{}c@{}}API-Call\\ (System-Call)\end{tabular}        & \begin{tabular}[c]{@{}c@{}}GetCurrentProcessId()\\ CreateToolhelp32Snap()\\ Process32Next()\\ NtQueryInfor.Proc.()\\ FindWindow()\end{tabular} & \cite{branco2012scientific, AntiDebugApiCallSystemCall}                & \begin{tabular}[c]{@{}c@{}}Modify results\\ via EPT-Hook (hiding process)\end{tabular}                                              \\
\multicolumn{1}{c|}{}                                              & PEB Field                                                               & \begin{tabular}[c]{@{}c@{}}IsDebuggerPresent()\\ NtGlobalFlags()\end{tabular}                                                                             & \cite{AntiDebugPeb}               & \begin{tabular}[c]{@{}c@{}}HyperDbg is not detectable \\ by default\end{tabular}                                                    \\
\multicolumn{1}{c|}{}                                              & Heap Structure                                                          & \begin{tabular}[c]{@{}c@{}}HEAP.Flags\\ HEAP.ForceFlags\end{tabular}                                                                                      & \cite{kimdefeating}               & \begin{tabular}[c]{@{}c@{}}HyperDbg is not detectable\\ by default\end{tabular}                                                     \\
\multicolumn{1}{c|}{}                                              & \#BP Detection                                                          & \begin{tabular}[c]{@{}c@{}}Find BP (0xCC) inst.	\\ Read DR (Debug Register)\end{tabular}                                                    & \cite{AntiDebugBpDetection}               & \begin{tabular}[c]{@{}c@{}}!dr to modify and disable\\ unwanted BPs\end{tabular}                                            \\
\multicolumn{1}{c|}{}                                              & \begin{tabular}[c]{@{}c@{}}Timing \\ Measurement\end{tabular}           & \begin{tabular}[c]{@{}c@{}}GetTickCount(),\\ QueryPerf.Counter,\\ GetLocalTime()\end{tabular}                                       & \cite{AntiDebugTimingMeasurement}               & \begin{tabular}[c]{@{}c@{}}EPT-Hook\\ Modification of results\end{tabular}                                                          \\
\multicolumn{1}{c|}{}                                              & Trap-Interrupt                                                          & \begin{tabular}[c]{@{}c@{}}Instruction Prefix,\\ INT 3, 0x2D,\\ Interrupt 0x41\end{tabular}                                                               & \cite{AntiDebugTrapInterrupt}               & \begin{tabular}[c]{@{}c@{}}Set Exception bitmap in\\ VMCS\end{tabular}                                                              \\
\multicolumn{1}{c|}{}                                              & \begin{tabular}[c]{@{}c@{}}Control Flow\\ Manupulation\end{tabular}     & \begin{tabular}[c]{@{}c@{}}NtSuspendThread(),\\ NtSetInf.Thread(),\\ CreateThread()\end{tabular}                                                   & \cite{AntiDebugControlFlowManupulation}               & \begin{tabular}[c]{@{}c@{}}HyperDbg not detectable\\ by default\end{tabular}                                                     \\ \hline
  \multicolumn{1}{c|}{\multirow{6}{*}{{\rotatebox[origin=c]{90}{\centering \textit{Anti-VM/Hypervisor/Emulation}}}}} & CPU Instructions                                                        & \begin{tabular}[c]{@{}c@{}}CPUID forces a VM-exit\\ certain info in VM\end{tabular}                                                          & \cite{dimva2016}               & \begin{tabular}[c]{@{}c@{}}VM-exit (CPUID result\\  modification)\end{tabular}                                                      \\
\multicolumn{1}{c|}{}                                              & \begin{tabular}[c]{@{}c@{}}Protection Model\\ Instructions\end{tabular} & \begin{tabular}[c]{@{}c@{}}SIDT, SLDT,  SGDT\\ STR, SMSW\end{tabular}                                                                                     & \cite{branco2012scientific}               & \begin{tabular}[c]{@{}c@{}}VM-exit (emulation and \\ modification)\end{tabular}                                                     \\
\multicolumn{1}{c|}{}                                              & \begin{tabular}[c]{@{}c@{}}Architectural\\ Delta-Timing\end{tabular}    & \begin{tabular}[c]{@{}c@{}}RDTSC+CPUID+RDTSC\\ RDTSC(P)+RDTSC(P)\end{tabular}                                                                       & \cite{dimva2016}               & \begin{tabular}[c]{@{}c@{}}HyperDbg Trans. Mode\\ (!hide command)\end{tabular}                                                 \\
\multicolumn{1}{c|}{}                                              & In/Out Instructions                                                     & Magic I/O port in VMware                                                                                                                                  & \cite{AntiDebugInOutInstructions}               & \begin{tabular}[c]{@{}c@{}}VM-exit handled\\ (I/O bitmap)\end{tabular}                                                  \\
\multicolumn{1}{c|}{}                                              & Invalid MSR Access                                                      & \begin{tabular}[c]{@{}c@{}}Invalid MSR issues \\ General Protection (\#GP) \end{tabular}                                                       & \cite{zhang2015using}               & \begin{tabular}[c]{@{}c@{}}Emulate !msrread/!msrwrite\\  command\end{tabular}                                                 \\
\multicolumn{1}{c|}{}                                              & Exception Handling                                                      & \begin{tabular}[c]{@{}c@{}}Try-Catch\\ General Protection Excep.\\ (\#GP)\end{tabular}                                                                 & \cite{AntiDebugExceptionHandling}               & \begin{tabular}[c]{@{}c@{}}Handled by default\\ Inject  routine into\\ user-mode\end{tabular} \\ \hline
\end{tabular}%
}
\label{tab:antilist}

\end{table}


\subsection{Timing Transparency in \textsc{HyperDbg}} \label{timingtransparency}
\textsc{HyperDbg}'s transparent mode offers a solution for hiding the virtualization timing leakage by identifying VM-detecting sequences and replacing the timing values with those of a non-virtualized system. To the best of our knowledge, \textsc{HyperDbg} is the first debugger to offer a practical means to modify timing footprints used by analyzer software to detect virtualized environments. By using a statistical model of the execution time, extensive timing profiling is executed prior to the launch of the VMM module in order to emulate timestamps as close to the normal operating condition of the guest OS as possible.


\lstset{numbers=left, numberstyle=\tiny, language=[x86masm]Assembler,  escapeinside={\%*}{*)},          
}
\begin{lstlisting}[caption={The timing measurement code by forcing VM-exit}, label={lst:exemplo1}, style=trans]
 %*\color{magenta}rdtscp *)  ; get the current time clock 
 %*\color{magenta} cpuid *)  ;Execute a serialization instruction (VM-exit)
 %*\color{magenta}rdtscp *)  ; Delta Timing
\end{lstlisting}

\begin{figure}[t]
  \begin{subfigure}[b]{0.48\columnwidth}
    \includegraphics[width=\linewidth]{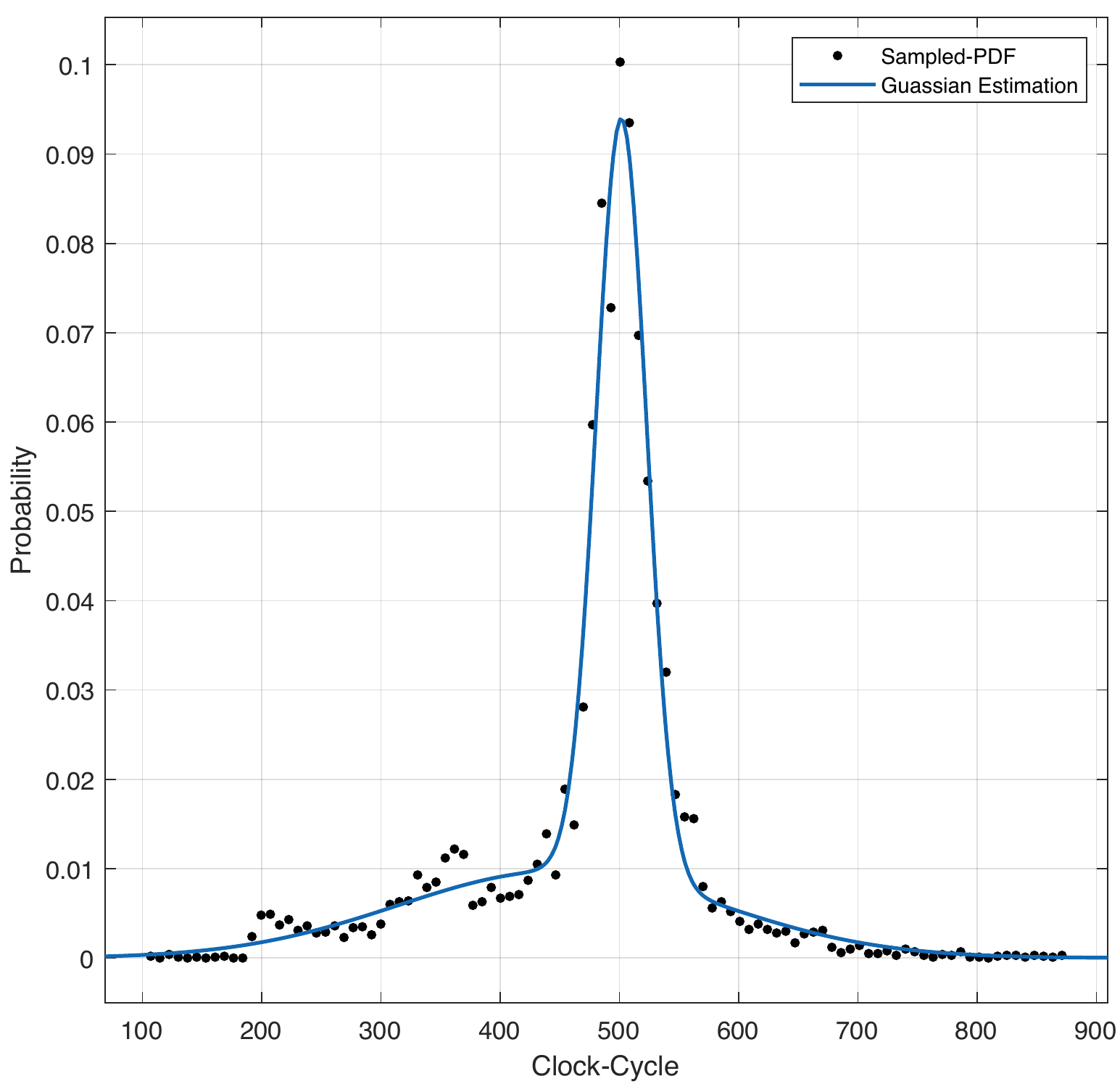}
    \caption{}
    \label{withouthyperdbg}
  \end{subfigure}
  \hfill 
  \begin{subfigure}[b]{0.48\columnwidth}
    \includegraphics[width=\linewidth]{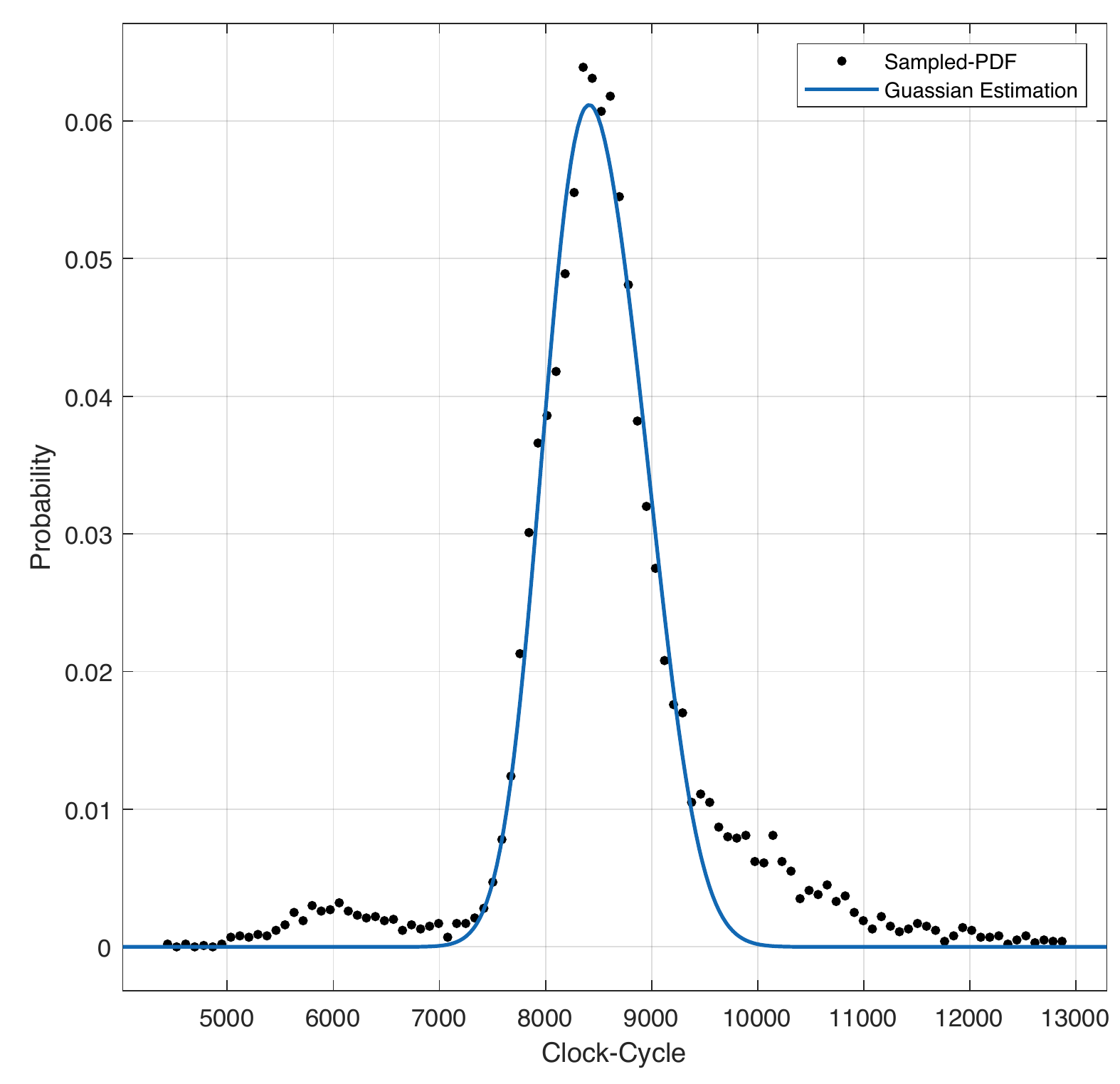}
    \caption{}
   \label{withhyperdbg}
  \end{subfigure}
  \caption{ PDF distribution of timing measurement for deactivated \textsc{HyperDbg} (a),  with activated \textsc{HyperDbg} (b)  }
  \label{fig:timing}
\end{figure}

We use a two-term Gaussian Distribution as a regressor function, as our experiments indicate that it can be a good fit for modeling the execution times of such nature. Figure \ref{fig:timing} shows the Probability Distribution Function (PDF) of our measurements by running 10k executions of the sequence in Listing \ref{lst:exemplo1}, with and without \textsc{HyperDbg} enabled. These values can be derived and the statistical parameters can be recorded for emulation purposes.



\begin{figure}[t] 
\centering
\includegraphics[width=0.8\linewidth]{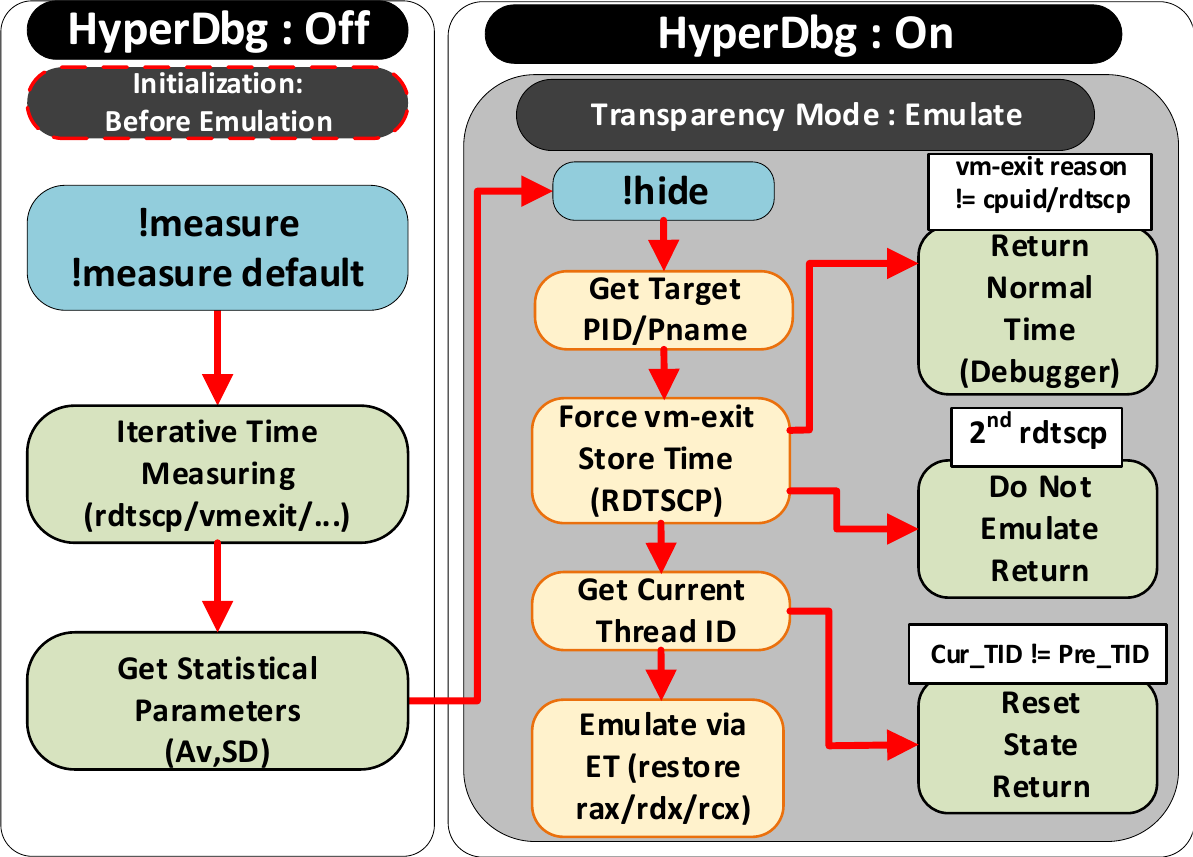}
\caption{State Diagram Process of \textit{rdtsc/rdtscp} emulation by \textsc{HyperDbg}}
\label{rdtc}
\end{figure}

\textsc{HyperDbg} currently enables covering the VM timing leakage by providing two methods that are: 1) Adjusting the MSR register that keeps track of the CPU's time which is referred as \texttt{IA32\_TIME}\\\texttt{\_STAMP\_COUNTER}, and 2) Emulating the results of \textit{RDTSC} and \textit{RDTSCP} instructions that provide the means for user-level applications for accessing the CPU timing values.  Each of these methods comes with its own set of advantages and setbacks. The former approach does not require a VM-exit for its function, which simplifies the solution and allows for covering more complex VM-detection sequences, but might increase the instability and interfere with the normal functionality of the system as other applications also rely on this mechanism for their timing measurements. In contrast, the latter method (Figure \ref{rdtc}) does not cause any interference with the inner workings of the system, but requiring a VM-exit adds a layer of complexity, as emulating timing values expected by an examiner program armed with sophisticated patterns for VM-detection would require extra considerations.

The transparency function can be used on a process or a list of executables, as global emulation of timing instructions would most likely disturb primary functionalities of the system (our experiments show disturbances in the screen driver, as well as audio output performance when a global emulation is implemented).

\subsection{Alternative Timing Attack Methods}

In addition to the methods explained in \ref{timingtransparency}, \textsc{HyperDbg} can impede various forms of timing attacks used by malware to expose the presence of a debugger. Since \textsc{HyperDbg} operates at the hypervisor level, it is possible to use \textsc{HyperDbg} to safeguard against other timing attacks that utilize alternative timing resources, such as shared ticks and Hardware Performance Monitor Counters (RDPMC), both of which have built-in support to defend against in \textsc{HyperDbg}'s Transparent Mode (See Table \ref{tab:events} in Appendix \ref{appen:A}). Additionally, while timing-thread attacks are one of the more challenging attacks to defend against, it is possible to detect such attempts using any known timing resources using \textsc{HyperDbg}. \textsc{HyperDbg} is actively adding methods that safeguard against new techniques used by malware for debugger detection.

\section{Evaluation}\label{sec:evaluation}

In this section, we thoroughly evaluate \textsc{HyperDbg}'s  transparency and performance in different scenarios. 
\subsection{Transparency Evaluation}
\label{sec:transeval}

We evaluate the transparency mode of \textsc{HyperDbg} using two of the best-known tools that offers stress-testing for anti-debugging and protection methods, pafish \cite{pafish} and al-khaser \cite{alkhaser}.

In accordance with our expectations, the first method, which involves updating the \texttt{IA32\_TIME\_STAMP\_COUNTER}, interferes with the primary functions of the system and causes screen flickering during our experiments. Regardless, the second method (emulation) was able to successfully pass these tools when enabling emulation for the anti-debugging testing software. As an extension of our transparency analysis, we separately evaluate \textsc{HyperDbg} against common anti-debugging methods and commercial off-the-shelf packers/protectors.

\subsubsection{Evaluation Configuration}

In our experiments, we analyzed 10,853 malware samples in different categories derived from a malware database \cite{vxunderground}. Each of these malware samples is executed in \textsc{HyperDbg}’s normal and transparent mode in the VMI Mode as well as \textit{x64dbg} (user-mode debugger) and \textit{WinDbg} (kernel-mode debugger) in Microsoft Windows 10 20H1 for comparison. We employ a client/server paradigm to distribute the samples among the client systems that execute the binaries and record logs from the executions. Each client asks the server for its according sample over a simple HTTP application. We use two approaches to restore the system: a rebootless, Barebox-based method \cite{kirat2011barebox} and an automated system restoration method based on Windows System Restore.

\textbf{Barebox-based Approach} We first attempt to run the malware under a setup that aims to rebootlessly restore the system using a set of Barebox-based methods to decrease the restoration time and improve performance. After fetching the malware from the server, the client globally disables interrupts and continues the main thread on a single core. While this can impose some performance penalties and communication issues with external devices, by removing the context switch to other processes, it simplifies the system restoration process and allows for observation of the effects of malware on the system exclusively, as it prevents unnecessary modifications to the structures unrelated to the subject binary.

Next, the automation program loads the binary and locates its entry point by capturing the page fault that follows in the initialization phase of the execution of a binary in Windows. A hardware breakpoint is then applied in this address (entry-point), which allows \textsc{HyperDbg} to be notified at the beginning of the program’s execution, once Windows is done with the initialization process of executing the binary. \textsc{HyperDbg} uses this to trigger the mechanism used for rebootless restoration of the contents of the memory.

To increase the performance of the testing process by limiting the restoration only to the modified pages, we make a snapshot of the clean installation of Windows by disabling writing on pages, which is accomplished by clearing every \textit{write} bit on EPT pages. This results in an EPT violation for a write request. We use the handler of this violation to make a clone of the pages that the malware attempts to write and then release the write lock to allow the binary to continue with its normal execution. While at the beginning of the execution, this can decrease the performance and sometimes make the system unresponsive, the performance improves as the execution progresses. To mitigate any resulting artifacts, \textsc{HyperDbg} is set to run in Transparent Mode.

Once the execution is finished, and the logs are saved using the VMX-root Mode Compatible Message Tracing Mechanism (Section \ref{messagetrace}), the interrupts are re-enabled, and our master program updates the order of runs and proceeds to restore the system. To restore the information on the volatile memory, we replace the pages that were cloned during the execution process back to their original state, while for disk writes, we use Shadow copy~\cite{ShadowCopy} combined with a mini-filter driver that monitors the modification/creation/deletion of the memory/registry to restore the disk to its initial state from the clean installation. This process takes around 7 minutes per malware sample.

If the system cannot be restored with this approach, we resort to the second approach, which requires a reboot and is based on Windows Restore Point. These cases include restoration of the system for bare metal executions, rootkit/bootkit samples, and cases where the execution encounters an error. A malware is classified as rootkit/bootkit using the tags provided by the malware providers and by detecting any attempt to load drivers. The success of the execution is measured using a Win32 API call \cite{GetExitCodeProcess}.

\textbf{Windows System Restore Approach} As an alternative approach, we rely on the Windows System Restore functionality. 
As there is no command-line tool for restoring snapshots, we implement a small tool to use the GUI tool automatically.
This method takes about 25-35 minutes to restore the system, based on the state of the system and the modifications made by the malware.

Overall, the testing process takes about 468 hours (168 hours testing malware with \textsc{HyperDbg} and 100 hours each for testing bare-metal, \textit{x64dbg}, and \textit{WinDbg}). We use 10 systems for this test.

\subsubsection{Evaluation by Anti-Debugging, -VM, and -Hypervisor}
 Table \ref{tab:antilist} describes the common anti-debugging and anti-virtualization methods \cite{afianian2019malware,zhang2016towards}, and \textsc{HyperDbg}'s countermeasure to impede detection. Each of these methods is applied separately in \textsc{HyperDbg}'s \textit{Debugger mode} and activates the suitable countermeasure to verify the transparency of the proposed debugger. 
 Furthermore, for an end to end transparency analysis, all the mitigation techniques are activated. 
We employ a combination of rebootless and reboot-based approaches for system restoration after the execution of each malware sample. 

We observe that a relatively large percentage of the samples detect the debugging environment in \textit{WinDbg} and \textit{x64dbg} and change their behavior accordingly to conceal their malicious behavior.  Considering \textit{WinDbg} is the baseline debugger, Figure \ref{fig:malwarerate} reports the percentage of successfully executed malware samples where the debugger is attached.  For this experiment, we measure the success rate of the execution of malware samples by carefully logging the syscall sequence in the target system by hooking the syscalls (changing IA32\_LSTAR). As shown in Figure \ref{fig:malwarerate}, \textsc{HyperDbg}'s Transparent mode increases the transparency surface by 22\% compared to \textit{WinDbg}, executing malware samples in all four categories while remaining undetected. This is due to the fact that \textsc{HyperDbg} operates at the hypervisor-level, which minimizes footprints that anti-debugging/VM methods in malware use to detect a debugging environment.  We manually investigated malware samples that detect the presence of \textsc{HyperDbg} and reverse engineered the binaries to inspect their inner-workings. We could attribute the detection of \textsc{HyperDbg} to two main factors. Employment of hypervisor-specific techniques that lead to the non-successful execution of malware in already virtualized environments, and utilization of methods that detect the absence of PatchGuard, or Driver Signature Enforcement (DSE) in the system. These methods can be countered by adding support for nested virtualization (cf. \ref{subsec:futureworks}) and obtaining a valid driver signature. However, this does not imply a claim of full invisibility upon the addition of said improvements. For example, \textsc{HyperDbg} is a new and open-source tool and it is possible for malware producers to focus on the stealthiness methods employed in this tool and find new methods to counter those efforts to reveal its presence.



\begin{figure}[t] 
 \centering
    \includegraphics[width=0.8\linewidth]{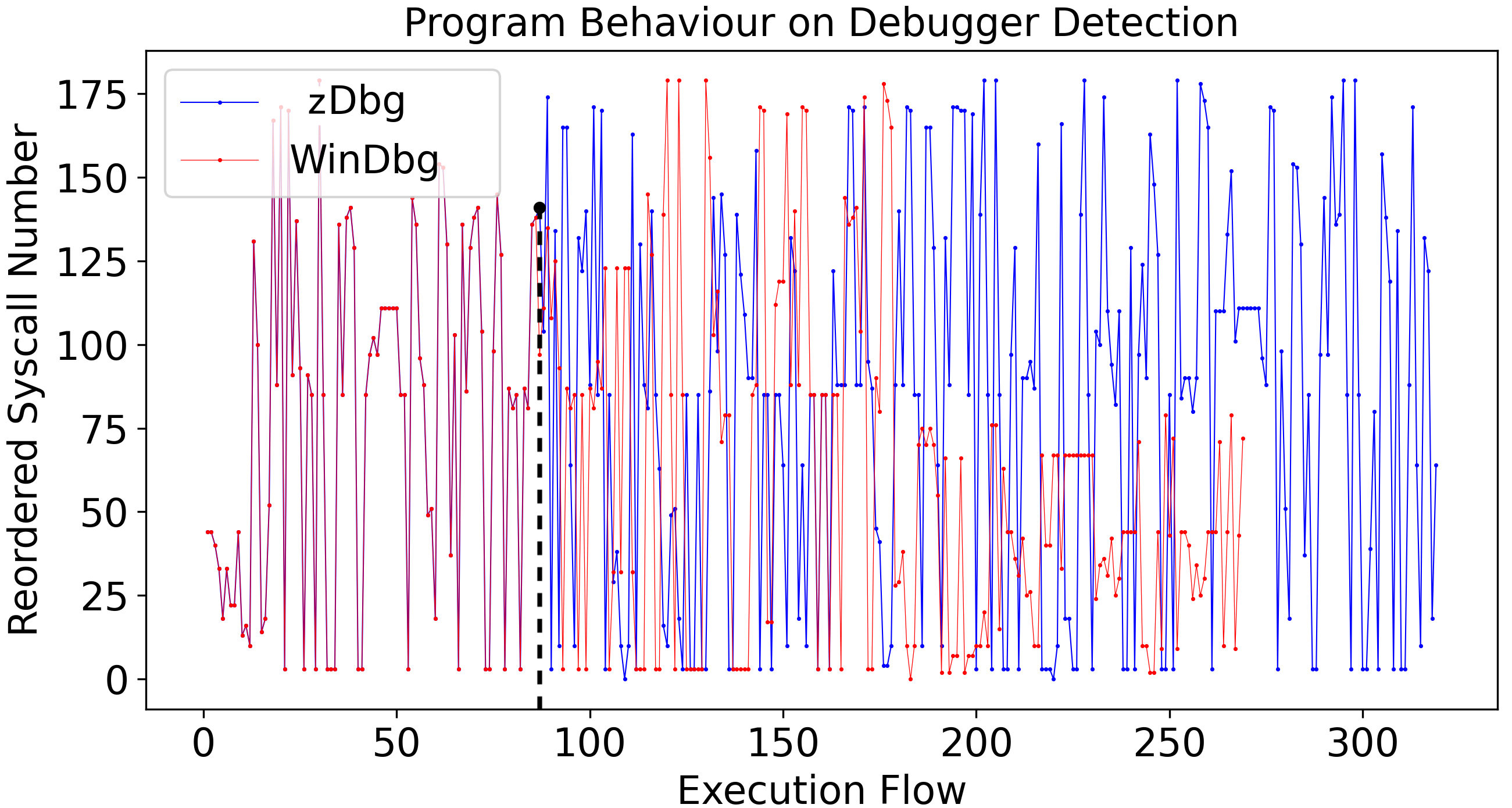}
\caption{Syscalls executed in a malware using \textsc{HyperDbg} and \textsc{WinDbg}}
\label{syscallmalware}
\end{figure}

\subsubsection{Syscall Malware Analysis}
\label{sec:sysmalware}
\begin{figure*}[t]
    \centering
     \begin{subfigure}[b]{0.30\textwidth}
\centering
\begin{tikzpicture}[scale=0.8]
\centering
\pgfmathsetmacro{\NPlotsFirstAx}{3}
\pgfmathsetmacro{\NPlotsSecondAx}{2}
\pgfmathsetmacro{\SumPlots}{\NPlotsFirstAx+\NPlotsSecondAx+2}
\pgfmathsetmacro{\WdFirst}{\NPlotsFirstAx/\SumPlots}
\pgfmathsetmacro{\WdSecond}{\NPlotsSecondAx/\SumPlots}
\begin{groupplot}[
       group style={
       group name=plot,
       group size=1 by 1,
       ylabels at=edge left,
       horizontal sep=0pt,
       vertical sep=0pt,
       /pgf/bar width=5pt,
       },
ylabel={ Executed Samples (\%)},
major x tick style=transparent,
ybar= 0.2pt,
ymin=-10,
ymax=40,
x=0.25\textwidth,
x axis line style={opacity=0},
x tick label style={rotate=0, anchor=center},
xticklabel style={yshift=-3mm,xshift={ifthenelse(\ticknum==2,0,0)}}, 
xtick=data,
ytick={-10,0,10,20,30,40},
xticklabels={Trojan, Rootkit, Spyware, Worm},
ymajorgrids=true,
grid style=dashed,
nodes near coords,
scale only axis,
enlarge x limits = {abs=1cm},
point meta=explicit symbolic,
cycle list={
  draw=black, line width = .1mm,fill=RYB1,fill opacity=1,nodes near coords style={black}\\
  draw=black, line width = .1mm,fill=RYB2,fill opacity=1,nodes near coords style={black}\\
  draw=black, line width = .1mm,fill=RYB3,fill opacity=1,nodes near coords style={black}\\
    draw=black, line width = .1mm,fill=RYB4,fill opacity=1,nodes near coords style={black}\\
  },
legend columns=-1,
height=0.5\textwidth,
width=1.6\textwidth,
 legend columns=-1,
        legend style={draw=none, /tikz/every even column/.append style={column sep=5pt},at={(axis cs:4.5,55)}},
        legend image code/.code={%
             \draw[#1] (0cm,-0.1cm) rectangle (0.2cm,0.1cm);
        }
]

\nextgroupplot
\addplot  coordinates {
  (1,-1.8)
  (2,-6.8)
   (3,-2.5)
  (4,-5.4)
};

  \addplot  coordinates {
  (1,23.2)
  (2,18.5)
   (3,25.62)
  (4,26.4)
};
  \addplot  coordinates {
  (1,26.1)
  (2,22.0)
   (3,30.92)
  (4,30.39)
};

\addplot[red,line legend,sharp plot,
    update limits=false] coordinates {(-2mm,0) 
    (4mm,0)};

\legend{x64dbg,  HDbg, HDbg(Tran),WinDbg}
\end{groupplot}


\end{tikzpicture}
\captionsetup{justification=centering}

    \caption{\scriptsize{Rate of successful executed malware samples in debuggers}}
        \label{fig:malwarerate}
     \end{subfigure}
     \begin{subfigure}[b]{0.3\textwidth}
  \centering

\begin{tikzpicture}[scale=0.8]

\pgfmathsetmacro{\NPlotsFirstAx}{3}
\pgfmathsetmacro{\NPlotsSecondAx}{2}
\pgfmathsetmacro{\SumPlots}{\NPlotsFirstAx+\NPlotsSecondAx+2}
\pgfmathsetmacro{\WdFirst}{\NPlotsFirstAx/\SumPlots}
\pgfmathsetmacro{\WdSecond}{\NPlotsSecondAx/\SumPlots}
\begin{groupplot}[
      group style={
      group name=plot,
      group size=1 by 1,
      ylabels at= edge right,
      horizontal sep=0pt,
      vertical sep=0pt,
      /pgf/bar width=7pt,
      },
major x tick style=transparent,
ybar= 0.3pt,
ymin=0,
ymax=12,
x=0.20\textwidth,
ytick={0,2,4,6,8,10,12},
x axis line style={opacity=0},
x tick label style={rotate=0, anchor=center},
xticklabel style={yshift=-2mm,xshift={ifthenelse(\ticknum==2,0,0)}}, 
xtick=data,
xticklabels={Trojan, Rootkit, Spyware, Worm},
ymajorgrids=true,
grid style=dashed,
nodes near coords,
scale only axis,
enlarge x limits = {abs=1cm},
point meta=explicit symbolic,
cycle list={
  draw=black, line width = .1mm,fill=RYB1,fill opacity=1,nodes near coords style={black}\\
  draw=black, line width = .1mm,fill=RYB2,fill opacity=1,nodes near coords style={black}\\
  draw=black, line width = .1mm,fill=RYB3,fill opacity=1,nodes near coords style={black}\\
    draw=black, line width = .1mm,fill=RYB4,fill opacity=1,nodes near coords style={black}\\
  },
legend columns=-1,
height=0.6\textwidth,
width=1.2\textwidth,
 legend columns=-1,
        legend style={draw=none, /tikz/every even column/.append style={column sep=5pt},at={(axis cs:4.8,15)}},
        legend image code/.code={%
             \draw[#1] (0cm,-0.1cm) rectangle (0.2cm,0.1cm);
        }
]

\nextgroupplot 
\addplot  coordinates {
  (1,10.7)
  (2,11.67)
  (3,10.47)
    (4,9.1)

};
  
\addplot  coordinates {
  (1,4.5)
  (2,4.7)
  (3,4.3)
  (4,4.1)
};

  \addplot  coordinates {
  (1,3.3)
  (2,3.6)
  (3,2.9)
    (4,3.3)

};

\legend{ Anti-VM/HV,Anti-Debug,Both}
\end{groupplot}


\end{tikzpicture}

\captionsetup{justification=centering}

    \caption{\scriptsize{Anti-Debug/VM/HV methods on Table \ref{tab:antilist} using \textsc{HyperDbg}}}
        \label{fig:antirate}

     \end{subfigure}
\begin{subfigure}[b]{0.3\textwidth}
\centering

\begin{tikzpicture}[scale=0.8]

\centering

\pgfmathsetmacro{\NPlotsFirstAx}{3}
\pgfmathsetmacro{\NPlotsSecondAx}{2}
\pgfmathsetmacro{\SumPlots}{\NPlotsFirstAx+\NPlotsSecondAx+2}
\pgfmathsetmacro{\WdFirst}{\NPlotsFirstAx/\SumPlots}
\pgfmathsetmacro{\WdSecond}{\NPlotsSecondAx/\SumPlots}
\begin{groupplot}[
       group style={
       group name=plot,
       group size=1 by 1,
       ylabels at=edge left,
       horizontal sep=0pt,
       vertical sep=0pt,
       /pgf/bar width=7pt,
       },
ylabel={Average No of Syscalls},
major x tick style=transparent,
ybar= 0.2pt,
ymin=0,
ymax=1000,
x=0.25\textwidth,
x axis line style={opacity=0},
x tick label style={rotate=0, anchor=center},
xticklabel style={yshift=-2mm,xshift={ifthenelse(\ticknum==2,0,0)}}, 
xtick=data,
xticklabels={Torjan, Rootkit, Spyware, Worm},
ymajorgrids=true,
grid style=dashed,
nodes near coords,
scale only axis,
enlarge x limits = {abs=1cm},
point meta=explicit symbolic,
cycle list={
  draw=black, line width = .1mm,fill=RYB1,fill opacity=1,nodes near coords style={black}\\
  draw=black, line width = .1mm,fill=RYB2,fill opacity=1,nodes near coords style={black}\\
  draw=black, line width = .1mm,fill=RYB3,fill opacity=1,nodes near coords style={black}\\
    draw=black, line width = .1mm,fill=RYB4,fill opacity=1,nodes near coords style={black}\\
  },
legend columns=-1,
height=0.5\textwidth,
width=1.6\textwidth,
 legend columns=-1,
        legend style={draw=none, /tikz/every even column/.append style={column sep=5pt},at={(axis cs:4.1,1350)}},
        legend image code/.code={%
             \draw[#1] (0cm,-0.1cm) rectangle (0.2cm,0.1cm);
        }
]

\nextgroupplot 
\addplot  coordinates {
  (1,658)
  (2,436)
   (3,862)
    (4,598)

};
  
\addplot  coordinates {
  (1,614)
  (2,387)
   (3,742)
   (4,491)
};

  \addplot  coordinates {
  (1,636)
  (2,395)
   (3,792)
    (4,512)

};

\legend{ \textsc{HyperDbg}, x64dbg, WinDbg}
\end{groupplot}


\end{tikzpicture}

\captionsetup{justification=centering}

    \caption{\scriptsize{Averaged number of executed Syscalls in malware samples in debuggers}}
        \label{fig:malwaresyscalnum}
     \end{subfigure}
        \caption{Evaluation results of malware execution experiments}
        \label{fig:three graphs}
\end{figure*}
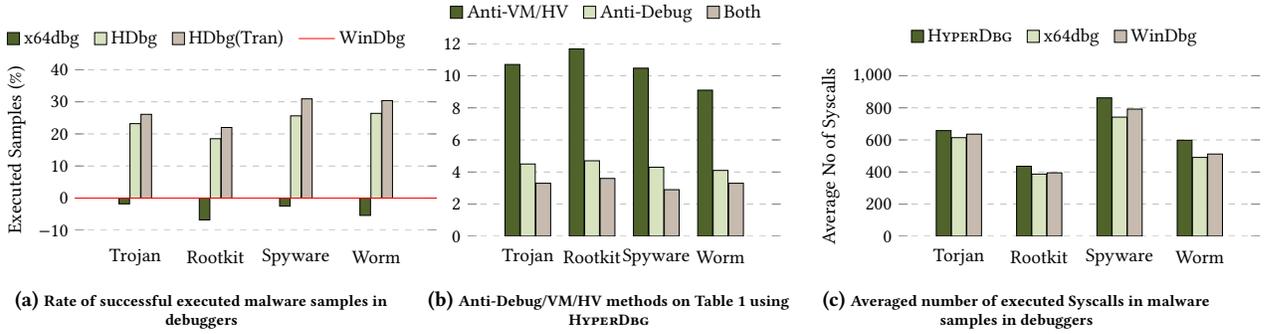

The transparent mode of this debugger offers the advantage of stealthily monitoring malware's execution. Figure \ref{syscallmalware} shows the syscall execution flow of a sample malware on \textsc{HyperDbg}. For high-level comparison it is possible to extract execution flow and divergence point of evasive malware here. One can execute the malware on a bare-metal system with no debugging present rather than \textsc{HyperDbg}'s transparent mode for monitoring purposes. We attach a kernel-mode debugger (\textit{WinDbg}) to the malware, execute the malware, and record the syscall execution flow. We use a simple script code using \textsc{HyperDbg}'s script engine to trace the \textit{SYSCALL}s in the execution flow. (See Appendix \ref{scriptenginecodesnippet})
As Figure \ref{syscallmalware} depicts, the execution flow of the malware does not follow a similar behavior in the different environments. As a simple analysis, we could come to the conclusion that this specific malware uses anti-debugging methods. To conceal its malicious intention, an entirely different (and most likely safe) execution path is chosen within the malware code when a debugger is detected. The same approach is used to measure whether the malware samples are running or not, as shown in Figure \ref{fig:malwarerate}.


\subsubsection{Evaluation by Packers and Protectors Testing}

\begin{table}[b]
\caption{Evaluation and comparison of \textsc{HyperDbg} for integrated software via packers/protectors }
\resizebox{1\columnwidth}{!}{%
\begin{tabular}{ccccccc}
\hline
\multicolumn{1}{l}{{Packer/Protector}} & \multicolumn{1}{l}{{File Type}} & \multicolumn{1}{l}{{WinDbg}} & \multicolumn{1}{l}{{x64dbg}} & \multicolumn{1}{l}{{Ollydbg }} & \multicolumn{1}{l}{{\textbf{HyperDbg}}} & {\begin{tabular}[c]{@{}c@{}}\textbf{HyperDbg's}\\  \textbf{Trans. Mode}\end{tabular}} \\ \hline
ASPack.V2.42                                  & PE32                                   & Error                               & \cmark                                   & \cmark                                             & \cmark                                     & \cmark                                                                               \\
Enigma.V4.30                                  & PE64                                   & \xmark                                  & \xmark                                  & N/A                                     & \xmark                                    & \cmark                                                                               \\
Net\_Crypto.V5                                & PE32                                   & \xmark                                  & \xmark                                  & \xmark                                            & \textbf{\cmark }                                   & \textbf{\cmark  }                                                                             \\
Obsidium.V1.7                                 & PE64                                   & \xmark                                  & \xmark                                  & N/A                                     & \cmark                                     & \cmark                                                                               \\
Themida.V3.0.4                                & PE64                                   & \xmark                                  & \xmark                                  & N/A                                     & \cmark                                     & \cmark                                                                               \\
Upx.V3.96                                     & PE64                                   & \cmark                                   & \cmark                                   & \cmark                                             & \cmark                                     & \cmark                                                                               \\
Vmprotect.V.2.13                              & PE64                                   & \xmark                                  & \xmark                                  & N/A                                     & \xmark                                    & \cmark                                                                               \\
MEW11.V1.2                                    & PE32                                   & \cmark                                   & \cmark                                   & \cmark                                             & \cmark                                     & \cmark                                                                               \\
Pecompact.V3.11                               & PE32                                   & \cmark                                   & \cmark                                   & \cmark                                             & \cmark                                     & \cmark                                                                               \\
PELock.V2.0                                   & PE32                                   & \xmark                                  & \xmark                                  & \xmark                                            & \cmark                                     & \cmark                                                                               \\
Petite.V2.4                                   & PE32                                   & Error                               & \cmark                                   & \cmark                                             & \cmark                                     & \cmark                                                                               \\
TeLock.V0.98                                  & PE32                                   & \cmark                                   & \cmark                                   & \cmark                                             & \cmark                                     & \cmark                                                                               \\
YodaCrypter.V1.02                             & PE32                                   & \xmark                                  & \xmark                                  & \xmark                                            & \cmark                                     & \cmark                                                                               \\ \hline
\end{tabular}%
}
\label{packertable}
\end{table}

We test \textsc{HyperDbg} with binaries processed with packers and protectors. These binaries are tested on different debuggers as well as \textsc{HyperDbg} in both regular debugging and transparent mode debugging. Table \ref{packertable} shows the results of attaching and debugging these protected binaries.

\subsection{Performance Evaluation}
In terms of performance, we analyze \textsc{HyperDbg} in three debugging scenarios which are discussed in the following. For our performance evaluations, we used a machine with an Intel core i7-6820HQ with 16GB of main memory, running Windows 10 20H1.
\subsubsection{Performance Analysis of Scenario 1: Step-in}
Single stepping is one of the most fundamental functionalities of a debugger that has been carefully optimized in \textsc{HyperDbg} to become as fast as possible. To evaluate the performance of this functionality, we considered $n=100$ sets of 65,536 predefined instructions (a particular application) to evaluate the performance.  \textsc{HyperDbg} was able to instrument the instruction sets on average in 6 minutes and 51 seconds ($\mu=411$ seconds) with the standard deviation of $\sigma=28.3$ seconds. It took \textit{WinDbg} on average $1,221$ seconds with the standard deviation of $\sigma=118.4$ seconds to perform the same function on the same instructions sets. Thus,  \textsc{HyperDbg} takes 2.97 less time on average to execute the same analysis compared to \textit{WinDbg}'s instrumenting. 
For a fair comparison with WinDbg, 
we also used Windows SDK's \textit{KDNET} as a transport for \textsc{HyperDbg}'s Ethernet-based communication. 
In this scenario, \textsc{HyperDbg} is 2.85 times faster.

\subsubsection{Performance Analysis of Scenario 2: Conditional Breakpoints}

An essential part of analyzing binaries are conditional breakpoints, which have been attempted to be implemented robustly and considerably faster than almost all of the currently available debuggers in \textsc{HyperDbg}. 

To evaluate the performance of \textsc{HyperDbg} on conditional breakpoints, we set one on the frequently used \textit{nt!ExAllocatePoolWi-} {thTag} function, checking whether the \textit{RAX} register contains a specific value. 
As performance metric, we count the number of times the condition was checked within 5 minutes both for \textsc{HyperDbg} and \textit{WinDbg}. 
To mitigate timing noise on our results, we repeat the experiments for $n=50$ times. As the baseline of the performance, \textit{WinDbg} checks on average 6,941 conditions. At the same time, \textsc{HyperDbg} checks on average 9,153,731 conditions with its classic implementation of EPT Hook and checks on average 23,214,792 conditions with its detours style EPT Hook. 

These result show that \textsc{HyperDbg}'s script engine achieves 1,319 and 3,345 average fold speedups compared to \textit{WinDbg} in classic EPT Hook and detours EPT Hooks, respectively.

This significant speed gain comes from the fact that based on the design, \textsc{HyperDbg} checks and evaluates scripts directly in the kernel and VMX-root mode and does not need the assistance of the user-mode for this end. Thus, in contrast to \textit{WinDbg}, nothing is transferred to the debugger during the execution of the script.

This difference is also visible in the system's overall performance during the execution of the benchmarks on the debugger. In \textit{WinDbg}, the system slows down to the point that it seems the system has come to a halt since not even the most basic computations, such as cursor movements are properly processed. While in \textsc{HyperDbg}'s case, even though the performance of the system is still slow, it's usable. Therefore, other tasks could still be normally performed on the system, which allows alteration and addition of new conditional breakpoints while the test is performed.

\subsubsection{Performance Analysis of Scenario 3: Analyzing Syscalls}

Setting breakpoints on syscalls is another scenario that can be used for evaluation of the performance of \textsc{HyperDbg}.

Generally speaking, it is not possible to set a breakpoint on syscall-handler routines in other debuggers like \textit{WinDbg}. However, it is possible to trace system calls by setting breakpoints on functions responsible for dispatching the SYSCALL numbers. 
In \textsc{HyperDbg}, it is possible to set breakpoints on syscall-handler routines and to emulate system calls.
For the performance evaluation, we perform $n=50$ experiments each lasting 300 seconds. \textit{WinDbg} executes 2,559 syscalls, while at the same time  \textsc{HyperDbg} executes 5,166,430.
Hence, \textsc{HyperDbg} is on average $\sim$2018x times more efficient than \textit{WinDbg} in tracing syscall routines.


\section{Applications}\label{sec:applications}
With the privileged access level and the newly-presented APIs, \textsc{HyperDbg} can be used in many applications.

\subsection{Debugging Devices}
\textsc{HyperDbg} supports the general functionality of any other debuggers, i.e., pausing and stepping through the instructions, read/write on memory, read and modify registers, and putting breakpoints anywhere in the program. Plus, it has many other creative events to ease the debugging process.








One of the unique capabilities of \textsc{HyperDbg} is its ability to debug the communications of the system with external devices. The user can monitor each x86 I/O port separately for port mapped I/O (PMIO) devices and use EPT to monitor Memory Mapped I/O (MMIO) devices. Since I/O instructions and EPT modifications are treated as events in \textsc{HyperDbg}, the user is able to monitor the executions of \textit{IN} and \textit{OUT} instructions and create separate logs.  Moreover, it is also possible to modify the registers in the script engine and therefore, delivering the modified values to the operating system.
In addition to debugging Port Mapped I/O and Memory Mapped I/O, \textsc{HyperDbg} is also capable of notifying the user about the interrupts from external devices. For example, \textsc{HyperDbg} can be configured to intercept any particular interrupt from an external device (e.g., a PS/2 keyboard) and allow the user to halt Windows to investigate the device in case it occurs, or simply ignore the interrupt and allow the operating system to continue normally. 

\subsection{Fuzzing}
One of the main problems of kernel fuzzing is the fact that every invalid value causes a kernel error and thus a BSOD. \textsc{HyperDbg} can avoid these errors by handling them even before the OS is notified and help fuzzing  (e.g., by measuring the code coverage).
As \textsc{HyperDbg} resides on a more privileged ring than the kernel, it can intercept code-level exceptions that lead to the crash of the system (application/OS) (e.g., page faults or division by zero) and discard the crash before calling the OS error handling routines.

\textsc{HyperDbg}'s script engine provides the possibility to execute brute-force tests for a target program using simple scripts. For instance, one can re-execute a target function arbitrarily often, each time with different parameters.

The proposed instrumentation step-in procedure in \textsc{HyperDbg} forces the system to only run the specific process without switching to other processes. Consequently, the CPU only executes the targeted codes during the fuzzing process and returns the program flow to the initial state of fuzzing if any crash appears. Using the script engine, it is then possible to prepare the CPU for the next stage of fuzzing with new parameters in an entirely automated mechanism. This method results in a fine-grained approach to fuzz both user and kernel programs with high-performance execution.

\subsection{Malware Analysis}
As another essential application, \textsc{HyperDbg} features a transparent debugging tool that can be used for evasive malware analysis. Given the unique toolset of  \textsc{HyperDbg}, online malware analysis is armed with a high-performance run-time script-engine, which makes the process effective and substantially faster. We describe a simple and transparent syscall malware analysis using \textsc{HyperDbg} in Section \ref{sec:sysmalware}. In the following, we survey the applicability of \textsc{HyperDbg} in a Windows vulnerability.

\subsubsection{Analysis of a Vulnerability: A Case Study}

 During our experiments, we rediscover a full-kernel mode Bootkit known as \textit{Pitou} \cite{payloadpitou}, to which the latest Windows versions are still vulnerable. We briefly describe \textit{Pitou} as a case study analyzed by \textsc{HyperDbg}.

Pitou is able to attack the victim system by bypassing the user access control and performing privilege escalation, which enables it to infect the Master Boot Record (MBR). This allows it to inject its kernel payload at the time of Windows startup without facing any resistance from Kernel Mode Code Signing (KMCS) policy. Pitou is then able to take control of the lowest level components of the OS (e.g., Windows network driver - NDIS) and utilize VM-level code that is not executable natively on Windows to obfuscate itself from conventional disassemblers, which makes it much more difficult to analyze it. To the date of writing, it is still able to infect the systems running the latest version of Microsoft Windows with a 0-day local privilege escalation.

Pitou also employs advanced anti-debugging and anti-sandboxing techniques that look for any traces of the execution in a non-native execution environment by performing inspections on Windows registry, kernel modules, disk devices, BIOS memory, and measurement of CPU ticks using RDTSC. These methods have been shown to be updated by the creator of the malware over time. In our tests, the malware detects the debugger environment with some of the most well-known and widely-used debuggers like \textit{WinDbg}, \textit{x64dbg}, and \textit{Ollydbg}\cite{OllyDbg}. It deviates from its normal behavior on every other debugger. However, with \textsc{HyperDbg}'s transparent mode,  we successfully execute the malware and perform an extensive dynamic analysis to reverse-engineer its execution flow.

\subsubsection{Digital Forensics \& Incident Response (DFIR)}
\textsc{HyperDbg} can be used extensively in the DFIR to detect signs of attacks. For instance, the script engine of \textsc{HyperDbg} can be utilized for developing a pre-built plugin to monitor the top abused APIs/syscalls under user-specified conditions and on any subset of the processes (e.g., critical system processes only), allowing the inspector to adjust between the conciseness and thoroughness of the logs based on their preference. Additionally, \textsc{HyperDbg} is capable of classifying the APIs into different categories of attacks (e.g., code injection, keylogging, or discovery) and transmitting the results over TCP/Named Pipe/File using Event Forwarding.


\subsubsection{Attempt to Exploit Detection}
\textsc{HyperDbg} can be used to detect many exploitation techniques. Often, exploits modify a special structure as the final payload, such as the token of a process~\cite{accessman,karvandi2022tsx}.

In the above example, \textsc{HyperDbg} can be used to monitor TOKEN structure and detect any access (or more precisely, any write) to this structure. After that, this abnormal behavior can be traced back to reach the initial phase of exploit and reveal its method.



\subsection{Software Performance Analysis}
\textsc{HyperDbg} can be used for performance and security analysis in software development and testing. For example, the highly optimized methods available in \textsc{HyperDbg} can be utilized for intercepting events such as page faults, with marginally superior performance compared to alternative means and methods used in an user-mode analysis tool \cite{hines2009post,bangert2013page,azimi2005online}.

\textsc{HyperDbg} can detect page faults in both the operating system and applications. 
In previous works, Shadow Paging, Page Tracking, and Pseudo-paging methods were used to detect page-faults~\cite{hines2009post}. Detecting page-faults is beneficial in the evaluation of applications that opt to improve their performance by minimizing the frequency of page faults. \textsc{HyperDbg} can detect page faults by exploiting exception bitmaps and providing it as an event. Using this method, \textsc{HyperDbg} can provide the exact address of fault area (CR2) for further investigations. This method is transparent to the operating system and does not change its semantics.



\section{Related Work}\label{sec:relatedwork}

Developing a debugger and low-level software analyzer has been regarded as a crucial topic for the computer community due to its impact and applicability in a wide range of scientific research and industrial products. The implications can be generally categorized into two main groups: 1) Hardware-level malware analysis and 2) System isolation, monitoring, and sandboxing.

Over the past decade, many researchers have proposed several debugging methods based on the ring -1 (sub-OS level) infrastructure to address these issues. However, in terms of transparency level, applicability, performance, and generalization, these tools fail to present a suitable solution for the community. \textsc{HyperDbg} as an open-sourced and general hardware-assisted debugger that aims to provide researchers and computer engineers with a tool to help resolve the aforementioned issues. 

 \textbf{Hardware-level malware analysis.}
Malware developers have managed to develop many strategies and techniques that allow them to escape from almost every form of detection methodology, including virtualization, debugging, and emulation techniques. 
Anti-debugging and anti-virtualization techniques used in early malware \cite{chen2008towards} employ numerous evasion methods to hide or reduce malicious activities. 
These anti-detection methods are analyzed comprehensively in the recent study by Galloro et al.~\cite{galloro2022systematical} where over 92 classes of evasive techniques executed by modern malware. 

Furthermore, hardware-based artifacts such as processor's cache actualization \cite{plumerault2021dbi}, scheduling leakage in simultaneous multithreading (SMT) \cite{lusky2021sandbox}, as well as timing side effects \cite{oyama2019does} of monitoring facilities can be observed by evasive malware.

As thoroughly discussed by Garfinkel et al.~\cite{garfinkel2007compatibility}, achieving full transparency against malware running in a virtualized environment is extremely challenging. Considering all of the issues, previous work proposed frameworks such as Apate \cite{shi2017hiding} to hide debugging procedures from malware. Likewise, other work proposes resilient malware detectors against evasive malware using hardware features~\cite{islam2021efficient,tian2021mdchd}. Leon et al.~\cite{leon2021hypervisor} study the possibility of utilizing hypervisors to detect, deactivate and analyze evasive malware by employing low-level processor features.

Unlike previous solutions, which merely focus on transparency rather than functionality, our method in \textsc{HyperDbg} to approach malware analysis provides a richly equipped debugging facility by pushing the deployment of more complex functionalities deeper into the hypervisor. This approach not only provides transparency but gains significant performance, as well as rich functionality all together in a singular framework making \textsc{HyperDbg} applicable for real-world malware analysis.

 \textbf{System isolation, tracing, and sandboxing.} Due to the increasing complexity of the malware evasion techniques, researchers have recently evolved the environment from VM-based sandboxes such as CWSandbox \cite{willems2007toward} and Cuckoo sandbox \cite{ccocko2020} to Bare-Metal sandboxes like BareBox \cite{kirat2011barebox}, and BareCloud \cite{kirat2014barecloud} to minimize the leakage of the virtualization environment. Pioneered by Ether \cite{dinaburg2008ether} as the first hypervisor-based analyzer with more transparency level, Malt \cite{zhang2016towards} and Ninja \cite{ning2017ninja} target Intel's SMM and Arm's TrustZone to present hardware-level debugging and process tracing as well as sandboxing primarily aiming at malware debugging. Although transparent to some level, all these works provide simple functionalities and low-speed tracing, making them unsuitable for deep and dynamic code analysis. \textsc{HyperDbg} addresses these shortcomings by providing real-time user-specified debugging functionalities using VMX-based script-engine. Furthermore, even though the hardware overhead is negligible for most previous solutions, the total debugging execution flow is prolonged due to the continuous ring transportation to perform dynamic code analysis. This drawback is fundamentally solved in \textsc{HyperDbg} 's design.

Recently, researchers employ newer hardware-based features (e.g. Intel-PT) for low-level hypervisor fuzzing \cite{schumilo2021nyx,schumilo2020hyper}, kernel failure reverse debugging \cite{ge2020reverse} as well as machine learning approaches \cite{arakelyan2021bin2vec}, to discover vulnerabilities and bugs.  Similar ideas are deployed for embedded systems arming application tracing \cite{du2020hart}, debugging\cite{ning2021revisiting,ning2019understanding},  unpacking\cite{xue2021happer} on Arm processors. 

Though designed as debugger, \textsc{HyperDbg} delivers high-level transparency for low-level sandboxing  and isolation. Moreover, its architectural design and VMX-enabled script engine provide an accurate and fast process tracking of arbitrary binaries.

\textbf{Feature comparison among existing debuggers}
A Comprehensive feature comparison is given in Table \ref{tab:com} in Appendix \ref{app:d}.

\section{Discussions and Limitations}

\textbf{Transparency} Several timing attacks for detecting \textsc{HyperDbg} are impeded. Still, by relying on external timing resources such as NTP, a binary can leverage measurement methods beyond the domain of the local system to detect \textsc{HyperDbg}'s timing adjustments. Nonetheless, these methods for debugger detection can potentially be mitigated in a case-specific analysis of a binary; however, the development of a general solution for this set of approaches is considered out of scope in this paper.


\textbf{Stability} Direct emulation of system-wide general-purpose mechanisms (such as modifying MSRs, cf. Section \ref{timingtransparency}) might  interfere with the
normal functionality of other applications that use timing measurements. 
Thus, we recommend using the second method (e.g., emulation of RDTSC and RDTSCP instructions) for already up and running test-bed environments. 
Furthermore, the Barebox-based restoration process requires manual restorations in some scenarios where Windows uses Asynchronous Procedure Calls (APC) as the inter-processor communication is disabled in such test-cases \cite{APCweb}.
We plan to develop a general testing framework and release it publicly as part of the \textsc{HyperDbg} project.

\textbf{Performance}
Although faster than similar debuggers, our experiments show cases where \textsc{HyperDbg} faces slowdowns du to an excessive number of VM-exits. 
To minimize the overhead, unnecessary VM-exits can be reduced, and emulation of the system at the early stage of VM-exits can be avoided to improve the overall performance of the system. 
For instance, the user might specify a specific core to apply the events.



\textbf{Future Works} \label{subsec:futureworks} In future releases, we intend to make a UEFI-based module for \textsc{HyperDbg}, which permits \textsc{HyperDbg} to run compatibly with Windows Virtualization-based Security (VBS). We also plan to add support for nested virtualization in \textsc{HyperDbg} to support the execution of nested guest VMs inside the virtualized environment of \textsc{HyperDbg}.

\section{Conclusion}\label{sec:conclusion}
With the expanding hardware support in modern processors, it is now more than ever crucial to employ hardware-assisted techniques in software debugging.
Common software debugging solutions rely on traditional OS-dependent APIs for code functionality analysis, vulnerabilities detection, and reverse engineering. Modern packed software and evasive malware employ sophisticated anti-debugging methods to hide their primary functionalities and withstand reverse engineering attempts on the extensively used debugging solutions.  This paper presents \textsc{HyperDbg}, an open-source hypervisor-level debugging tool with transparency and performance in mind. \textsc{HyperDbg} exploits Intel VT-x and Intel EPT to present multiple new debugging modules, useful for fuzzing, malware analysis, and reverse engineering.
We propose a novel VMX-level script engine in \textsc{HyperDbg}'s core  which gives an unmatched debugging performance useful for software fuzzing as user-mode to kernel-mode (and vice versa) transfer is entirely avoided. Our evaluation shows a high level of stealth code analysis against malware classes and unprecedented performance in terms of debugging functionality among other available kernel debuggers.
Finally, \textsc{HyperDbg} is designed modular and scalable for convenient usage in both academia and industry.

\bibliographystyle{ACM-Reference-Format}
\bibliography{ref}

\section*{Appendices}

\appendix
\label{Appendices}
\section{Event Commands in \textsc{HyperDbg}}
\label{appen:A}
\begin{table}[H]
\caption{The list of the supported events in \textsc{HyperDbg}}
\label{tab:events}
\centering
\resizebox{0.8\columnwidth}{!}{%
\begin{tabular}{|c|c|}
\hline
\color[HTML]{00379C}  !epthook  & Classic EPT hook                                                                                                                        \\ \hline
\color[HTML]{00379C}  !epthook2  & EPT hook with detours hooking                                                                                                           \\ \hline
\color[HTML]{00379C}  !syscall  & Hook execution of system-calls                                                                                                          \\ \hline
\color[HTML]{00379C}  !sysret   & Hook execution of sysret instruction                                                                                                    \\ \hline
\color[HTML]{00379C}  !monitor   & \begin{tabular}[c]{@{}c@{}}Monitors any access (Read/Write) \\ to a region of memory\end{tabular}                                       \\ \hline
\color[HTML]{00379C}  !cpuid     & \begin{tabular}[c]{@{}c@{}}System-wide CPUID instruction \\ execution detection\end{tabular}                                            \\ \hline
\color[HTML]{00379C}  !msrread   & \begin{tabular}[c]{@{}c@{}}System-wide RDMSR instruction \\ execution detection\end{tabular}                                            \\ \hline
\color[HTML]{00379C}  !msrwrite  & \begin{tabular}[c]{@{}c@{}}System-wide WRMSR instruction\\  execution detection\end{tabular}                                            \\ \hline
\color[HTML]{00379C}  !tsc       & \begin{tabular}[c]{@{}c@{}}System-wide RDTSC/RDTSCP\\  instructions execution detection\end{tabular}                                    \\ \hline
\color[HTML]{00379C}  !pmc       & \begin{tabular}[c]{@{}c@{}}System-wide RDPMC instruction\\  (performance counter) execution detection\end{tabular}                      \\ \hline
\color[HTML]{00379C}  !exception & \begin{tabular}[c]{@{}c@{}}Monitors and hooks first 32 entries\\  of Interrupt Descriptor Table (IDT)\end{tabular}                      \\ \hline
\color[HTML]{00379C}  !interrupt & \begin{tabular}[c]{@{}c@{}}Monitors and hooks external-interrupts 33 to \\ 256 entries of Interrupt Descriptor Table (IDT)\end{tabular} \\ \hline
\color[HTML]{00379C}  !dr        & \begin{tabular}[c]{@{}c@{}}Detects any reads or write \\ into hardware debug registers\end{tabular}                                     \\ \hline
\color[HTML]{00379C}  !ioin      & \begin{tabular}[c]{@{}c@{}}Monitors and ability to modify \\ I/O ports and IN instruction\end{tabular}                                  \\ \hline
\color[HTML]{00379C}  !ioout     & \begin{tabular}[c]{@{}c@{}}Monitors and ability to modify \\ I/O ports and OUT instruction\end{tabular}                                 \\ \hline
\color[HTML]{00379C}  !vmcall    & \begin{tabular}[c]{@{}c@{}}System-wide VMCALL instruction \\ (hypercalls) execution detection\end{tabular}                              \\ \hline
\end{tabular}%
}

\end{table}

\section{Process/Processor/Execution Mode Switch}
In this section, we describe the architecture related to switching between processes, processors, and different modes of execution in \textsc{HyperDbg}.

\subsection{Detecting Execution-mode Changes (Kernel-mode to User-mode)}
Detecting changes to the operating mode is performed via the same mechanism used in the \textit{i} command in \textsc{HyperDbg}. 
The proper method for implementing this functionality would be checking the CS register, fetching GDT, and checking the Long Mode flag. However, since the CS for \textit{wow64} and native code is set to a constant value across all versions of Windows, the CS register check is sufficient for the determination of a mode switch.

\subsection{Switching to a New Processor/Process}

\textsc{HyperDbg} uses a straightforward mechanism to switch between cores. Each core has its own spinlock to wait on VMX-root mode. By unlocking the spinlock assigned to the new core and setting the spinlock of the current core, it is possible to enter a waiting state. Consequently, \textsc{HyperDbg} calls the command handler from the new core, making the new core responsible for getting commands.
Note that \textsc{HyperDbg} is designed to have a single core for getting commands at any given time.
Moreover, switching to a new process is performed by monitoring changes to the \textit{CR3} register. Each time Windows changes the memory layout of any process, the \textit{CR3} is changed, and \textsc{HyperDbg} checks whether or not Windows has switched to the target process. If the memory layout is changed and the target process is now on the execution stage of the Windows, \textsc{HyperDbg} halts the debuggee again and waits for the commands from the debugger.



\subsection{Getting Debugging Events: \#BPs and \#DBs}

\textsc{HyperDbg} uses the exception bitmap of VMCS to get notified of breakpoints (\#BP) and Debug Breakpoints (\#DB) to halt the other cores. \textsc{HyperDbg} is the first debugger capable of being notified about the debugging event which means that \textsc{HyperDbg} is notified even earlier than the operating system. Consequently, we design the system not to notify user-mode application or kernel-mode (OS) entities regarding the debugging events. So, all the breakpoints events are handled by \textsc{HyperDbg}.

\subsection{Spinning on Spinlocks}
Spinning the cores in \textsc{HyperDbg} is considered as a primary technique in its functionalities. We study the challenges in this context. Suppose a function requires a spinlock (e.g. it is merely a buffer which is to be accessed) in a single-core processor. The function raises the IRQL to \textit{DISPATCH\_LEVEL}. Here, the Windows Scheduler can not interrupt the function until it releases the spinlock and lowers the IRQL to \textit{PASSIVE\_LEVEL} or \textit{APC\_LEVEL}. If during the execution of the function, a VM-exit occurs, the operation mode is moved into the VMX-root. (It can be interpreted that a VM-exit happens similar to a HIGH\_IRQL interrupt.)

In the case where a user accesses the buffer in the VMX-root mode, two scenarios are possible:
\begin{itemize}
    \item The first scenario is to wait on a spinlock that was previously acquired by a thread in the VMX non-root mode. In such scenario, a deadlock occurs and spins forever.
    \item Alternatively, it is also possible to enter the function without looking at the lock (while there is another thread that enters the function at the same time), which would result in a corrupted buffer and invalid data.
Windows also imposes another limitation, as cores must not wait on a spinlock when IRQL is higher than \textit{ DISPATCH\_LEVEL}. This lies in fact that Windows raises the IRQL to (\textit{DISPATCH\_LEVEL}) 2, when a spinlock is acquired. In this case, Windows performs the workload, releases the spinlock and lowers IRQL back afterwards.
\end{itemize}

Looking at corresponding Windows functions (e.g \textit{KeAcquireSpinLock} and \textit{KeReleaseSpinLock}), the IRQL arguments are given as input. Windows saves the current IRQL to the parameter supplied by the user in \textit{KeAcquireSpinLock}and then it raises the IRQL to \textit{DISPATCH\_LEVEL}. After the function is finished with the shared data, it calls \textit{KeReleaseSpinLock} and passes the old IRQL parameter to the function. Finally, it unsets the bit and restore the old IRQL (lowering the IRQL).



Unfortunately, Windows spinlocks employs IRQLs which do not make sense when VMX-root mode is in action. This makes it very complicated to use such functions in this mode. Hence, to implement spinlock for \textsc{HyperDbg} functionalities such as multi-core message tracing, we design a custom VMX-root compatible spinlock.

\subsection{MTFs Disadvantages}
 By setting the monitor trap flag, it is not necessarily guaranteed that the next instruction is the targeted instruction. In this case, if the upcoming instruction is a sudden interrupt from the CPU, the next targeted instruction in the debugging program would not be executed since the interrupt handler instructions are executed first. 
One way to address this issue is to set a \textit{VM-exit} on exceptions (Exception Bitmap) and external-interrupts. However, this resolution is not optimal as it might causes system inconsistency by blocking interrupts.
\textsc{HyperDbg} is able to resolve this issue using an instrumentation stepping process. 

The following Listing illustrates the set/unset  of MTF in an execution sequence.


\begin{lstlisting}[style=CStyle,
caption=MTF Set/Unset in an example execution sequence, label=listing:1]
/* Set the monitor trap flag */
void HvSetMonitorTrapFlag(BOOLEAN Set)
{
 unsigned long CpuBasedVmExecControls = 0;
 // Read the previous flag
 __vmx_vmread(CPU_BASED_VM_EXEC_CONTROL, &CpuBasedVmExecControls);
 if (Set) {
 CpuBasedVmExecControls |= CPU_BASED_MONITOR_TRAP_FLAG;
 }
 else {
 CpuBasedVmExecControls &= ~CPU_BASED_MONITOR_TRAP_FLAG;
 }
 // Set the new value 
 __vmx_vmwrite(CPU_BASED_VM_EXEC_CONTROL, CpuBasedVmExecControls);
}
\end{lstlisting}

\subsection{Debugger Pausing}

Typically, there are two scenarios in which the kernel debugger is paused. A breakpoint is triggered either by a break request from an event or the script engine. In this context, if the user is in the kernel-mode, a \textit{VMCALL} occurs and the future chain of events are handled accordingly. If the user is already in \textit{VMX-root} mode, other cores should be notified to prevent a system-level crash. Operating in VMX-root mode is similar to HIGH\_IRQL. In VMX-root mode, all the interrupts are masked because of RFLAGS' IF Bit.

The other scenario is upon the request from the user (for instance, an interruption by pressing CTRL+C). There, a packet is sent to pause the debugger. In this method, the debugger processes the packet in user mode, invoking an \textit{IOCTL}, executing a \textit{VMCALL} which transfers from the kernel-mode to the \textit{VMX-root}.

\subsection{Continuation a Single Core}

One of the exclusive features of \textsc{HyperDbg} is to keep execution (continuation) on one core while other cores are in a halt-state. We used this mechanism in our instrumentation step-in command to guarantee that no other cores (threads) get the chance to be executed. The fundamental basis of this mechanism is to ensure that the target core is not interrupted during the debugging.

There are two approaches that \textsc{HyperDbg} uses to prevent a target core from getting interrupted (e.g. by clock interrupt or keyboard interrupts).
\begin{itemize}
    \item First, the \textit{RFLAGS.IF} bit of the guest can be unset so the interrupts are masked. 
    \item Second, the \textit{PIN Based External-Interrupt Exiting bit}  can be set so all of the external interrupts would cause VM-exits; thus, allowing the interrupts to be simply ignored in the VM-exit handler
    
\end{itemize}

The former method is faster and avoids unnecessary VM-exits. However, for several considerations described in following, the second method is preferred in \textsc{HyperDbg}.

In contrast to the method described in the first approach, it is much safer not to change the guest's registers. As an example, if a page-fault, \textit{SYSCALL}, or an invalid operation such as division-by-zero occurs, the execution is directed to the kernel and guest's \textit{RFLAGS} are saved by the processor. Therefore, extra operations are required to locate the user-mode RFLAGS (search in stack for exceptions and in \textit{R11 Register} for \textit{SYSCALLs}), because the \textit{RFLAGS} that was previously saved in user-mode is with IF bit disabled. If this specific task is ignored, \textit{RFLAGS} are restored without checking for IF bit every time the guest continues and performs a context-switch. In this case, by unsetting this bit from hypervisor, the core becomes uninterruptible as the OS cannot get the execution again (e.g, using clock interrupt). Consequently, after a delayed bug check, Windows realizes the target core behaves abnormally and returns an error. Moreover, changing the guest's \textit{RFLAGS} is also incompatible with instructions like \textit{CLI} and \textit{STI}. More importantly, considering the side effects, the guest is able to detect the tampering of \textsc{HyperDbg} using \textit{PUSHF} function and check for \textit{IF} bit in RFLAGS.  

All of the issues investigated regarding \textit{RFLAGS} changing, in addition to the fact that using PIN Based External-Interrupt Exiting bit is completely transparent from the kernel-mode and user-mode, has lead us to employ the second method in \textsc{HyperDbg}.

\section{Full Descriptions of The Used Terms}
\label{appen:C}

\begin{table}[!h]
\caption{List of the used terms and their brief description. }
\label{tab:term}
\resizebox{1\columnwidth}{!}{%
\begin{tabular}{|cc|c|}
\hline
\multicolumn{2}{|c|}{Term}                                                                                                                                                & \multirow{2}{*}{Description}                                                                                                                                                                                \\ \cline{1-2}
\multicolumn{1}{|c|}{Abbreviation}                                                        & Full Form                                                                     &                                                                                                                                                                                                             \\ \hline
\multicolumn{1}{|c|}{\begin{tabular}[c]{@{}c@{}}VMX-root\\ VMX non-root\end{tabular}}     & \begin{tabular}[c]{@{}c@{}}Virtual Machine\\  Extensions\end{tabular}         & \begin{tabular}[c]{@{}c@{}}Intel's Modes of Operations:\\ A software on VMX-root mode\\ has higher privileges, has access to certain \\ instructions regardless of the privilege level\end{tabular}         \\ \hline
\multicolumn{1}{|c|}{VM-Entries}                                                          & -                                                                             & \begin{tabular}[c]{@{}c@{}}Transition From VMX-root to \\ VMX non-root\end{tabular}                                                                                                                         \\ \hline
\multicolumn{1}{|c|}{VM-Exits}                                                            & -                                                                             & \begin{tabular}[c]{@{}c@{}}Transition From VMX non-root to \\ VMX-root\end{tabular}                                                                                                                         \\ \hline
\multicolumn{1}{|c|}{VMCS}                                                                & \begin{tabular}[c]{@{}c@{}}Virtual Machine \\ Control Structure\end{tabular}  & \begin{tabular}[c]{@{}c@{}}A hardware-defined structure to control \\ the settings of each guest VM.\end{tabular}                                                                                           \\ \hline
\multicolumn{1}{|c|}{\begin{tabular}[c]{@{}c@{}}PB VM-\\ Execution Controls\end{tabular}} & \begin{tabular}[c]{@{}c@{}}Processor-Based \\ Execution Controls\end{tabular} & \begin{tabular}[c]{@{}c@{}}Part of VMCS which control features \\ and the vital attributes  of the hypervisors\end{tabular}                                                                                 \\ \hline
\multicolumn{1}{|c|}{MTF}                                                                 & Monitor Trap Flag                                                             & \begin{tabular}[c]{@{}c@{}}At the VMCS, permits operation of a processor \\ in single step mode in VMX-root.\\ following a VMRESUME, a single\\  instruction and then a VM-Exit occurs.\end{tabular}        \\ \hline
\multicolumn{1}{|c|}{-}                                                                   & Exception Bitmap                                                              & \begin{tabular}[c]{@{}c@{}}An 32-bit field in\\ the VMCS that controls the processor exceptions.\end{tabular}                                                                                               \\ \hline
\multicolumn{1}{|c|}{-}                                                                   & Event Injection                                                               & \begin{tabular}[c]{@{}c@{}} Events injected by Hypervisor (Interrupts, Exceptions, \\ NMIs, and SMIs) and these events  will be delivered\\ to the guest as if they’ve arrived normally\end{tabular} \\ \hline
\multicolumn{1}{|c|}{NMI}                                                                 & Non-Maskable Interrupts                                                       & \begin{tabular}[c]{@{}c@{}}A hardware-based\\ interrupt that cannot be ignored by the system.\end{tabular}                                                                                                  \\ \hline
\multicolumn{1}{|c|}{SYSCALL}                                                             & System Calls                                                                  & \begin{tabular}[c]{@{}c@{}}System calls are a programmatic way \\ that a user-mode application requests\\  a service from the operating system’s kernel.\end{tabular}                                       \\ \hline
\multicolumn{1}{|c|}{IRQL}                                                                & Interrupt Request Level                                                       & \begin{tabular}[c]{@{}c@{}}A hardware-independent mechanism\\ MS Windows uses to prioritize interrupts and code\\ that is running on the Processor.\end{tabular}                                         \\ \hline
\end{tabular}%
}
\end{table}

\subsection{VM-entries, VM-exits (VMX-root and VMX non-root)}

VT-x introduces two new modes of operations: VMX-root operation, and VMX non-root operation. A software running on VMX-root mode has higher privileges and has access to certain instructions that are not available in VMX non-root operation, regardless of the privilege level \cite{pratt2005xen}.
The core of \textsc{HyperDbg} runs mainly in the VMX-root mode, while guests (operating system's kernel, and applications) are executed in VMX non-root. 

With definition of these two modes of operation, VT-x consequently defines two new transitions: one being VM-entry, which is a transition from the root operation to guest non-root operation, and the other VM-exit, which performs the opposite. 

\subsection{Virtual Machine Control Structure (VMCS)}

To control the guest features, we have to set some properties in the Virtual Machine Control Structure (VMCS). The VMCS is a hardware-defined structure that controls the behavior and settings of each guest virtual machine (VM). Such a data structure exists for each VM in the memory and it is managed by the Virtual Machine Monitor (VMM). With every change of the execution context between different VMs, the VMCS is restored for the current VM, defining the state of the VM’s virtual processor and this way, VMM controls the guest software.

The VMCS consists of six logical groups:

\begin{itemize}
\item \textbf{Guest-State Area}: Processor state saved into the guest state area on VM-exits and loaded on VM-entries.
 \item \textbf{Host-State Area}: Processor state loaded from the host state area on VM-exits.
\item \textbf{VM-Execution Control Fields}: Fields controlling processor operation in VMX non-root operation.
\item \textbf{ VM-Exit Control Fields}: Fields that control VM-exits.
\item \textbf{ VM-Entry Control Fields}: Fields that control VM-entries.
\item \textbf{ VM-Exit Information Fields}: Read-only fields to receive information on VM-exits describing the cause and the nature of the VM-exit.
\end{itemize}

\subsection{Primary/Secondary Processor-Based VM-Execution Controls}

Primary Processor-Based VM-Execution Controls~\cite{primary_procbased2019intel}, a part of VMCS, along with Secondary Processor-Based VM-Execution Controls fields~\cite{primary_procbased2019intel,secondary_vmbased2019intel}, control features that can be altered using the VMWRITE instruction.

These fields control some of the vital attributes that determine the behavior of the hypervisors \cite{primary_procbased2019intel,secondary_vmbased2019intel} such as Monitor Trap Flags, or Enabling EPT.

\subsection{Monitor Trap Flag (MTF)}
The Monitor Trap Flag or MTF (located in the VMCS) is a feature provided by Intel that works exactly like the Trap Flag in \textit{RFLAGS}, which is additionally invisible to the guest. By setting this flag, following a VMRESUME, the processor is forced to execute a single instruction and then a VM-exit occurs.

\subsection{Exception Bitmap}
The exception bitmap is a 32-bit field in the VMCS that controls the exceptions (Exceptions are reserved by Intel on first 32 entries of Interrupt Descriptor Table).

When an exception occurs, its vector is used to select a bit in the exception bitmap. If the bit is 1, the exception will lead to a VM-exit. If the bit is 0, the exception is delivered normally through the IDT.

\subsection{Event Injection}
Hypervisors are able to inject events (Interrupts, Exceptions, NMIs, and SMIs) and these events are delivered to the guest as if they have arrived normally. Event injection is managed by special fields in VMCS.

\subsection{Non-Maskable Interrupts (NMI)}
An NMI is a hardware-based interrupt which typically cannot be ignored by the system. An NMI is usually triggered on hardware errors. These errors include non-recoverable internal system chipset errors, system memory corruptions such as parity and ECC errors, and data corruption detection on system and peripheral buses. Nevertheless, it is possible to mask specific NMIs  by employing special techniques.


Some OS-level programs use debugging NMIs to diagnose, analyze, and fuzz codes. In such cases an NMI can execute an interrupt handler transferring the control flow to a special monitor program. In this circumstance, the developer can monitor the memory and examine the internal state of the program at the instant of its interruption. This also allows the debugging or diagnosing of computers which appear hung.

On some systems, a computer software could drive an NMI through hardware and software debugging interfaces and system reset buttons. \textsc{HyperDbg} as a hypervisor-level software, employs NMI triggering extensively for kernel debugging.

\subsection{Syscalls}

Syscalls are a programmatic way that a user-mode application requests a service from the operating system's kernel. Syscalls provide an interface between a process and operating system and allow a user-level process to request a privileged service from the operating system.

In modern operating systems, syscalls are implemented using the SYSCALL and SYSRET instructions. These instructions rely on a set of Model Specific Registers (MSRs), namely \textit{IA32\_STAR}, \textit{IA32\_CSTAR}, and \textit{IA32\_LSTAR} MSR~\cite{pfoh2011nitro}.

This mechanism can be turned on and off by setting and unsetting the SCE flag in the Extended Feature Enable Register (EFER). Making use of either SYSCALL or SYSRET with the SCE flag not set results in an invalid opcode exception.

\subsection{Interrupt Request Level (IRQL)}

An Interrupt Request Level (IRQL) is a hardware-independent mechanism that Windows uses to prioritize interrupts and code that is running on the processors. Processes running at a higher IRQL will preempt a thread or interrupt running at a lower IRQL~\cite{IrqlMicrosoft}.

The below list shows different routines and the corresponding IRQL that these routines are running on:
\begin{itemize}

\item \textbf{DIRQL} : Interrupt Service Routines (ISRs) of hardware and external devices 
\item \textbf{DISPATCH\_LEVEL} : Scheduler, DPCs, and codes protected by a spinlock
\item \textbf{APC\_LEVEL} : Asynchronous Procedure Calls (APC) routines
\item \textbf{PASSIVE\_LEVEL} : User code, dispatch routines, and PnP routines
\end{itemize}

\section{Feature Comparison among Existing Debuggers}
\label{app:d}
Table \ref{tab:com} illustrates a  comparison among the existing debuggers and \textsc{HyperDbg}.
\begin{table*}[t]
\centering
\caption{A comprehensive comparison of \textsc{HyperDbg} with different debuggers}
\label{tab:com}
\resizebox{2\columnwidth}{!}{%
\tabcolsep=0.06cm
\begin{tabular}{c|c|c|c|c|c|c|c|c|c|c}

\multirow{2}{*}{Debugger}   & \multirow{2}{*}{\begin{tabular}[c]{@{}c@{}}Deployment\\  Level\end{tabular}} & \multicolumn{2}{c|}{Debugging Mode}  & \multirow{2}{*}{\begin{tabular}[c]{@{}c@{}}Transparency\\  Insurance\end{tabular}}                        & \multirow{2}{*}{\begin{tabular}[c]{@{}c@{}}Direct Deployment \\ (NO VM/Emulator)\end{tabular}} & \multirow{2}{*}{\begin{tabular}[c]{@{}c@{}}Source \\ Code Available\end{tabular}} & \multirow{2}{*}{\begin{tabular}[c]{@{}c@{}} Hooking\\\end{tabular}} & \multirow{2}{*}{Scripting}                                                             & \multirow{2}{*}{\begin{tabular}[c]{@{}c@{}}Custom Assembly\\  Execution\end{tabular}} & \multirow{2}{*}{\begin{tabular}[c]{@{}c@{}} Notable Features\\ (Currently Working Debuggers) \end{tabular}} \\ \cline{3-4}
                                                          &                                                                              & User Mode       & Kernel Mode       &                                                                                   &                                                                                                           &                                                                                                                                                                                     &                                                                                         &                                                                                        &                                                                                       &                                                                                     \\ \hline
\textbf{\textsc{HyperDbg}}                                                  & \begin{tabular}[c]{@{}c@{}}Hypervisor\\ (Ring -1)\end{tabular}               & \cmark             & \cmark                 & \begin{tabular}[c]{@{}c@{}}Hardware Assisted Methods\\ (e.g. TimeStamp Emulation \\ Transparency)\end{tabular} & \cmark                                                                                            & \cmark                                                                                 & \begin{tabular}[c]{@{}c@{}} EPT Hidden \\ Hooking\end{tabular}                                                                                   & \begin{tabular}[c]{@{}c@{}}Customized VMX\\ compatible \\ ScriptEngine\end{tabular}    & \cmark                                                                                   & \begin{tabular}[c]{@{}c@{}}Different subsystems\\ Fast script engine\\ I/O debugging \\ \end{tabular}       \\ 
\hline
Ghidra\cite{GHIDRA}                                                    & \begin{tabular}[c]{@{}c@{}}Application\\ (Ring 3)\end{tabular}               & \cmark             & \xmark              & \xmark                                                                                                      & \xmark                                                                                           & \cmark                                                                                 & \xmark                                                                                    & Python Scripting                                                                       & \xmark                                                                                  & \begin{tabular}[c]{@{}c@{}}Advanced Decompiler\\ Multi-platform\\ Multi-architecture support \\ \end{tabular}       \\ 
\hline
WinDbg\cite{windbg}                                                   & \begin{tabular}[c]{@{}c@{}}Operating \\ System\end{tabular}                  & \cmark             & \cmark                         & \xmark                                                                                                      & \cmark                                                                                            & \xmark                                                                                & \xmark                                                                                    & \begin{tabular}[c]{@{}c@{}}JavaScript\\ WinDbg Script\end{tabular}                     & \xmark                                                                                  & \begin{tabular}[c]{@{}c@{}}Javascript support\\ Advanced scripting language \\ \end{tabular}       \\
\hline

SoftIce\cite{softice}                                                   & \begin{tabular}[c]{@{}c@{}}Operating\\ System\end{tabular}                   & \xmark            & \cmark                         & \xmark                                                                                                      & \cmark                                                                                            & N/A                                                                               & \xmark                                                                                    & \xmark                                                                                   & \xmark                                                                                  & \begin{tabular}[c]{@{}c@{}}Local debugging with special GUI \\ \end{tabular}       \\
\hline
x64dbg\cite{x64}                                                    & \begin{tabular}[c]{@{}c@{}}Application\\ (Ring 3)\end{tabular}               & \cmark             & \xmark                        & \begin{tabular}[c]{@{}c@{}}Software-Based\\ ScyllaHide\end{tabular}                                       & \cmark                                                                                            & \cmark                                                                                 & \xmark                                                                                    & \begin{tabular}[c]{@{}c@{}}Customized \\ Scripting\end{tabular}                        & \xmark                                                                                  & \begin{tabular}[c]{@{}c@{}}Flexible and strong GUI \\ Lots of useful features \\ \end{tabular}       \\
\hline
Ollydbg\cite{OllyDbg}                                                   & \begin{tabular}[c]{@{}c@{}}Application\\ (Ring 3)\end{tabular}               & \cmark             & \xmark                        & \begin{tabular}[c]{@{}c@{}}Software-Based\\ ScyllaHide\end{tabular}                                       & \cmark                                                                                            & \xmark                                                                                & \xmark                                                                                    & ODBGScript                                                                             & \cmark                                                                                   & \begin{tabular}[c]{@{}c@{}}Stability \end{tabular}       \\
\hline
gdb \cite{GDB}                                                      & \begin{tabular}[c]{@{}c@{}}Operating\\ System\end{tabular}                   & \cmark             & \cmark                         & \xmark                                                                                                      & \cmark                                                                                            & \cmark                                                                                 & \xmark                                                                                    & \begin{tabular}[c]{@{}c@{}}Bash Scripting\\ Automation\end{tabular}                    & \xmark                                                                                  & \begin{tabular}[c]{@{}c@{}}Main Linux Debugger \\ Support multiple platforms \end{tabular}       \\
\hline
Malt\cite{zhang2016towards}                                                      & \begin{tabular}[c]{@{}c@{}}SMM \\ (Ring -2)\end{tabular}                     & N/A           & \cmark                  & \begin{tabular}[c]{@{}c@{}}SMM Level\\ Transparency\end{tabular}                                          & \cmark                                                                                            & N/A                                                                               & N/A                                                                                   & N/A                                                                                  & N/A                                                                                 & N/A                                                                                \\ 
\hline
BareBox\cite{kirat2011barebox}                                                   & \begin{tabular}[c]{@{}c@{}}Hypervisor\\ (Ring -1)\end{tabular}               & N/A           & \cmark                         & \begin{tabular}[c]{@{}c@{}}Meta OS (Bare Metal)\\ Transparency\end{tabular}                               & \cmark                                                                                            & N/A                                                                               & N/A                                                                                   & N/A                                                                                  & N/A                                                                                 & N/A                                                                                \\ 
\hline
V2E\cite{yan2012v2e}                                                      & \begin{tabular}[c]{@{}c@{}}Hypervisor\\ (Ring -1)\end{tabular}               & N/A           & \cmark                         & \begin{tabular}[c]{@{}c@{}}Hypervisor Level\\ Transparency\end{tabular}                                   & \xmark                                                                                           & N/A                                                                               & N/A                                                                                   & N/A                                                                                  & N/A                                                                                 & N/A                                                                                \\
\hline
Anubis\cite{mandl2009anubis}                                                    & \begin{tabular}[c]{@{}c@{}}Hypervisor\\ (Ring -1)\end{tabular}               & N/A           & \cmark                                                                                           & \begin{tabular}[c]{@{}c@{}}Limited Software-Based\\ Methods\end{tabular}                                  & \xmark                                                                                           & N/A                                                                               & N/A                                                                                   & N/A                                                                                  & N/A                                                                                 & N/A                                                                                \\ 
\hline
Virt-ICE\cite{quynh2010virt}                                                 & \begin{tabular}[c]{@{}c@{}}Hypervisor\\ (Ring -1)\end{tabular}               & N/A           & \cmark                         & \begin{tabular}[c]{@{}c@{}}Emulating Software-Based\\ Methods\end{tabular}                                & \xmark                                                                                           & N/A                                                                               & N/A                                                                                   & N/A                                                                                  & N/A                                                                                 & N/A                                                                                \\ 
\hline
\begin{tabular}[c]{@{}c@{}}HyperDbg (old)\\  (deprecated)\end{tabular} & \begin{tabular}[c]{@{}c@{}}Hypervisor\\ (Ring -1)\end{tabular}               & \cmark             & \cmark                         & N/A                                                                                                     & \xmark                                                                                           & \cmark                                                                                 & \xmark                                                                                    & \xmark                                                                                   & \xmark                                                                                  & N/A                                                                                \\ 
\hline
Ether\cite{dinaburg2008ether}                                                     & \begin{tabular}[c]{@{}c@{}}Hypervisor\\ (Ring -1)\end{tabular}               & \cmark             & \cmark                         & \begin{tabular}[c]{@{}c@{}}Hypervisor Level\\ Transparency\end{tabular}                                   & \xmark                                                                                           & N/A                                                                               & N/A                                                                                   & N/A                                                                                  & N/A                                                                                 & N/A                                                                                \\ 
\hline
VAMPiRE\cite{vasudevan2005stealth}                                                   & \begin{tabular}[c]{@{}c@{}}Hypervisor\\ (Ring -1)\end{tabular}               & N/A           & \cmark                         & \begin{tabular}[c]{@{}c@{}}Hypervisor Level\\ Transparency\end{tabular}                                   & \cmark                                                                                            & N/A                                                                               & N/A                                                                                   & N/A                                                                                  & N/A                                                                                 & N/A                                                                                \\ 
\hline
SPIDER\cite{deng2013spider}                                                   & \begin{tabular}[c]{@{}c@{}}Hypervisor\\ (Ring -1)\end{tabular}               & N/A           & \cmark                         & \begin{tabular}[c]{@{}c@{}}Hypervisor Level\\ Transparency\end{tabular}                                   & \xmark                                                                                           & N/A                                                                               & N/A                                                                                   & N/A                                                                                  & N/A                                                                                 & N/A                                                                                \\
\hline
IDAPro\cite{idapro}                                                   & \begin{tabular}[c]{@{}c@{}}Application\\ (Ring 3)\end{tabular}               & \cmark             & \xmark                        & \xmark                                                                                                      & \cmark                                                                                            & \xmark                                                                                & \xmark                                                                                    & \begin{tabular}[c]{@{}c@{}}Built-in \\ Scripting Engine \\ (IDC / Python)\end{tabular} & \xmark                                                                                   & \begin{tabular}[c]{@{}c@{}}Advanced decompiler \\ Multi-architecture \\ Multi-platform support \end{tabular}       \\
\end{tabular}%

}

\end{table*}
\section{HyperDbg's Script Engine}


\label{scriptenginecodesnippet}

\begin{lstlisting}[style=CStyle,
caption=Script engine examples, label=listing:4]
!syscall 0x55 pid 0x14c0 script {
    if(@rcx == 0x27 && @rdx == 0x47) {
        printf("Syscall triggered : %x in process id : %x\nThe third param : %llx\nThe fourth param : %llx\n", @rax, $pid, @r8, @r9);
        pause();    }
        }

\end{lstlisting}

In the above example, a syscall event is configured to trigger exclusively for syscalls specific to the process (pid = 0x14c0), which will execute the target script in VMX-root mode.

Considering that Windows uses fastcall calling convention for its syscalls, we know the registers stored in \textit{RCX}, \textit{RDX}, \textit{R8}, \textit{R9}, and stack. In the example script, it is checked if the first parameter to the syscall (\textit{RCX} register) is equal to 0x27 and the second parameter (\textit{RDX} register) is equal to 0x47. If these conditions are met, a message is printed, which generates a log from the 3rd (\textit{R8} register) and 4th (\textit{R9} register) parameters. At last, the \texttt{pause()} function is used to pause the debuggee and give the control to the debugger.

\section{Hardware Features on ARM and AMD}
\label{appen:amd}
ARM processors also contain virtualization extensions that provide hardware means for hypervisors to virtualize the CPU, permitting multiple OSes to be run on the same machine. ARM processors support the Second Level Address Translation (SLAT), which is known as Stage-2 page tables provided by a Stage-2 MMU.

Similarly, AMD supports virtualization through AMD-v technology. SLAT implementation in AMD processors is through the Rapid Virtualization Indexing (RVI), or Nested Page Tables (NPT) technology.

With the similar approach it is most likely possible to investigate and implement the same methodologies presented in this work for AMD or ARM-based hardware-assisted debuggers. 

\section{Application: Reverse Engineering}

\subsection{Automatic Symbol Reconstruction}
\label{appen:re}
One of the main goals of  \textsc{HyperDbg} is to provide a reverse engineering and dynamic binary analysing tool. Here, we describe a simple example of reverse-engineering an application. 
\textsc{HyperDbg} provides a functionality that maps a virtual memory to a C/C++ data type (e.g., enums, and structures). This is extremely useful, as OS-level PDB linkers can dynamically be translated to structures for testing and reverse engineering. For instance, a user can exactly detect the location of values in an specific function used by OS process and modify it accordingly. 
This is easily done using simple scripting.
The following listing shows how memory content of  target process can be easily delivered by structured representation. For instance, here \_PROCESS\_CREATION\_INFO is de-referenced and could be modified.

% ...

% ...

\begin{lstlisting}[caption={Converting PDB references to structures for reverse engineering}, label={lst:exemplo3}, style=trans]
 %*\color{magenta}kHyperDbg> *)  dt nt!_EPROCESS ffff948cc2393080
 _EPROCESS 
 ...
 %*\color{magenta} +0x05b7 *)  uint8_t PriorityClass : %*\color{magenta} 0x2 *)
 %*\color{magenta} +0x05b8 *)  void* SecurityPort : %*\color{magenta} (null) *)
 %*\color{magenta} +0x05c0 *) _SE_AUDIT_PROCESS_CREATION_INFO ... : %*\color{magenta} ffff948c`c25fa2a0 *)
 %*\color{magenta} +0x05c8 *) _LIST_ENTRY JobLinks ... : %*\color{magenta} [ 00000000`00000000 - 00000000`00000000 ] *)
\end{lstlisting}

\end{document}